
\input epsf
\def\blankline{\vskip \baselineskip}

\def\half{{\leavevmode\kern.1em\raise.5ex\hbox{\the\scriptfont0 1}\kern-.1em
/\kern-.1em\lower.25ex\hbox{\the\scriptfont0 2}}} 
\def\quarter{{\leavevmode\kern.1em\raise.5ex\hbox{\the\scriptfont0 1}\kern-.1em
/\kern-.1em\lower.25ex\hbox{\the\scriptfont0 4}}}

\def\spose#1{\hbox to 0pt{#1\hss}}
\def\ltsim{\mathrel{\spose{\lower 3pt\hbox{$\mathchar"218$}}
     \raise 2.0pt\hbox{$\mathchar"13C$}}}
\def\gtsim{\mathrel{\spose{\lower 3pt\hbox{$\mathchar"218$}}
     \raise 2.0pt\hbox{$\mathchar"13E$}}}
\def\gtlt{\mathrel{\spose{\lower 3pt\hbox{$\mathchar"13E$}}
     \raise 3pt\hbox{$\mathchar"13C$}}}

\def\today{\count99=\day
           \ifnum\count99>20 \count98=\day
                             \divide\count98 by 10
                             \multiply\count98 by 10
                             \advance\count99 by -\count98 \fi
           \number\day\ifcase\count99 th\or st\or nd\or rd\else th\fi
           ~\ifcase\month none\or January\or February\or March\or April\or
                  May\or June\or July\or August\or September\or October\or
                  November\or December\fi
           ~\number\year}

\newdimen\digitwidth
\setbox0=\hbox{0}
\digitwidth=\wd0

\def\etal{{\it et al.}}

\newcount\linespacingstep
\newcount\linespacing
\linespacingstep=1
\def\multiplelines{\linespacing=\linespacingstep
   \advance\linespacing by 3
   \multiply\normalbaselineskip by \linespacing
   \advance\normalbaselineskip by 2pt 
   \divide\normalbaselineskip by 3}

\font\fiverm=cmr5	\font\sixrm=cmr6	\font\sevenrm=cmr7
\font\eightrm=cmr8 	\font\ninerm=cmr9	\font\tenrm=cmr10
\font\twelverm=cmr12 	

\font\fivei=cmmi5 	\font\sixi=cmmi6 	\font\seveni=cmmi7
\font\eighti=cmmi8 	\font\ninei=cmmi9 	\font\teni=cmmi10
\font\twelvei=cmmi12

\font\fivesy=cmsy5  	\font\sixsy=cmsy6 	\font\sevensy=cmsy7
\font\eightsy=cmsy8 	\font\ninesy=cmsy9 	\font\tensy=cmsy10
\font\magnifiedtensy=cmsy10 at 12pt

\font\fivebf=cmbx5  	\font\sixbf=cmbx6 	\font\sevenbf=cmbx7
\font\eightbf=cmbx8 	\font\ninebf=cmbx9 	\font\tenbf=cmbx10
\font\twelvebf=cmbx12

	\font\eightit=cmti8 	\font\nineit=cmti9
\font\tenit=cmti10	\font\twelveit=cmti12

\font\eightsl=cmsl8 	\font\ninesl=cmsl9	\font\tensl=cmsl10
\font\twelvesl=cmsl12

\font\eighttt=cmtt8 	\font\ninett=cmtt9 	\font\tentt=cmtt10
\font\twelvett=cmtt12

\font\tenex=cmex10

\def\eightpoint{\def\rm{\fam0\eightrm}
\textfont0=\eightrm\scriptfont0=\sixrm\scriptscriptfont0=\fiverm
\textfont1=\eighti\scriptfont1=\sixi\scriptscriptfont1=\fivei
\textfont2=\eightsy\scriptfont2=\sixsy\scriptscriptfont2=\fivesy
\textfont3=\tenex\scriptfont3=\tenex\scriptscriptfont3=\tenex
\textfont\itfam=\eightit\def\it{\fam\itfam\eightit}
\textfont\slfam=\eightsl\def\sl{\fam\slfam\eightsl}
\textfont\ttfam=\eighttt\def\tt{\fam\ttfam\eighttt}
\textfont\bffam=\eightbf\scriptfont\bffam=\sixbf
\scriptscriptfont\bffam=\fivebf\def\bf{\fam\bffam\eightbf}
\normalbaselineskip=9pt
\ifnum\linespacingstep>1\multiplelines\fi
\setbox\strutbox=\hbox{\vrule height7pt depth2pt width0pt}
\let\sc=\sixrm\let\big=\eightbig\normalbaselines\rm}

\def\ninepoint{\def\rm{\fam0\ninerm}
\textfont0=\ninerm\scriptfont0=\sixrm\scriptscriptfont0=\fiverm
\textfont1=\ninei\scriptfont1=\sixi\scriptscriptfont1=\fivei
\textfont2=\ninesy\scriptfont2=\sixsy\scriptscriptfont2=\fivesy
\textfont3=\tenex\scriptfont3=\tenex\scriptscriptfont3=\tenex
\textfont\itfam=\nineit\def\it{\fam\itfam\nineit}
\textfont\slfam=\ninesl\def\sl{\fam\slfam\ninesl}
\textfont\ttfam=\ninett\def\tt{\fam\ttfam\ninett}
\textfont\bffam=\ninebf\scriptfont\bffam=\sixbf
\scriptscriptfont\bffam=\fivebf\def\bf{\fam\bffam\ninebf}
\normalbaselineskip=11pt
\ifnum\linespacingstep>1\multiplelines\fi
\setbox\strutbox=\hbox{\vrule height8pt depth3pt width0pt}
\let\sc=\sevenrm\let\big=\ninebig\normalbaselines\rm}

\def\tenpoint{\def\rm{\fam0\tenrm}
\textfont0=\tenrm\scriptfont0=\sevenrm\scriptscriptfont0=\fiverm%
\textfont1=\teni\scriptfont1=\seveni\scriptscriptfont1=\fivei%
\textfont2=\tensy\scriptfont2=\sevensy\scriptscriptfont2=\fivesy%
\textfont3=\tenex\scriptfont3=\tenex\scriptscriptfont3=\tenex%
\textfont\itfam=\tenit\def\it{\fam\itfam\tenit}%
\textfont\slfam=\tensl\def\sl{\fam\slfam\tensl}%
\textfont\ttfam=\tentt\def\tt{\fam\ttfam\tentt}%
\textfont\bffam=\tenbf\scriptfont\bffam=\sevenbf%
\scriptscriptfont\bffam=\fivebf\def\bf{\fam\bffam\tenbf}%
\normalbaselineskip=12pt%
\ifnum\linespacingstep>1\multiplelines\fi
\setbox\strutbox=\hbox{\vrule height8.5pt depth3.5pt width0pt}%
\let\sc=\eightrm\let\big=\tenbig\normalbaselines\rm}

\def\twelvepoint{\def\rm{\fam0\twelverm}
\textfont0=\twelverm\scriptfont0=\eightrm\scriptscriptfont0=\sixrm
\textfont1=\twelvei\scriptfont1=\eighti\scriptscriptfont1=\sixi
\textfont2=\magnifiedtensy\scriptfont2=\eightsy\scriptscriptfont2=\sixsy
\textfont3=\tenex\scriptfont3=\tenex\scriptscriptfont3=\tenex
\textfont\itfam=\twelveit\def\it{\fam\itfam\twelveit}
\textfont\slfam=\twelvesl\def\sl{\fam\slfam\twelvesl}
\textfont\ttfam=\twelvett\def\tt{\fam\ttfam\twelvett}
\textfont\bffam=\twelvebf\scriptfont\bffam=\eightbf
\scriptscriptfont\bffam=\sixbf\def\bf{\fam\bffam\twelvebf}
\tt 
\normalbaselineskip=14pt
\ifnum\linespacingstep>1\multiplelines\fi
\setbox\strutbox=\hbox{\vrule height10pt depth5pt width0pt}
\let\sc=\eightrm\let\big=\twelvebig\normalbaselines\rm}

\twelvepoint

\vsize=22.6 true cm \hsize=17 true cm 
\clubpenalty=5000
\widowpenalty=5000

\hoffset=0.cm
\voffset=0.cm
\hsize=16.cm
\vsize=21.cm

\font\headfont=cmti12
\font\titlefont=cmr12
\font\authorfont=cmcsc10 at 12pt
\font\affilfont=cmr8
\font\lonefont=cmr10
\font\ltwofont=cmti12

\def\sect#1{\par\goodbreak\bigskip
\centerline{\nextsect\lonefont\uppercase{#1}}\nobreak\vskip 3pt \nobreak}

\def\subsect#1{\par\goodbreak\medskip
\centerline{\nextsub\ltwofont~#1}}

\newcount\sectcount
\newcount\subcount
\newcount\ssubcount
\def\nextsect{\ifnum\sectcount>-1 \global\advance\sectcount by 1
        \number\sectcount. \global\subcount=0 \fi}
\def\nextsub{\global\advance\subcount by 1 \number\sectcount.\number\subcount.
        \global\ssubcount=0}
\def\nextssub{\global\advance\ssubcount by 1
\number\sectcount.\number\subcount.\number\ssubcount.}
\sectcount=0


\def\refs{\parskip=0pt\sect{References}
\parindent=0pt\everypar{\hangindent 1cm}
\ifnum\linespacingstep=1 \tenpoint \fi

\def\AAp{{A\&A,} }
\def\AApS{{A\&AS,} }
\def\AJ{{AJ,} }
\def\AnnRev{{ARA\&A,} }
\def\ApJ{{ApJ,} }
\def\ApJL{{ApJ,} }
\def\ApJS{{ApJS,} }
\def\ApSS{{Astrophys. Sp. Sci.,} }
\def\BAN{{Bull. astr. Inst. Netherlands,} }
\def\Cambridge{(Cambridge: Cambridge University Press)}
\def\JCP{{J. Comput. Phys.\/} }
\def\MNRAS{{MNRAS,} }
\def\Kluwer{(Dordrecht: Kluwer)}
\def\Messenger{{ESO Messenger} }
\def\Nature{{Nature\/} }
\def\Obs{{Observatory,} }
\def\PASJ{{PASJ,} }
\def\PASP{{PASP,} }
\def\PhD{{PhD thesis,} }
\def\PhilTrans{{Phil. Trans. R. Soc. Lond. A,} }
\def\PhysFl{{Phys. Fluids,} }
\def\PhysRep{{Phys. Reports} }
\def\Reidel{(Dordrecht: Reidel)}
\def\RMP{{Rev. Mod. Phys.} }
\def\RPP{{Rep. Prog. Phys.} }
\def\SovAst{{Sov. Astron.} }
\def\SovAstL{{Sov. Astron. Letters} }
\def\Vistas{{Vistas Astron.} }

}

\newcount\notenumber
\notenumber=1
\newcount\eqnumber
\eqnumber=1
\newcount\fignumber
\fignumber=1
\newbox\abstr
\def\chaphead{}
\def\oneskip{\vskip\baselineskip}

\def\note#1{\footnote{$^{\the\notenumber}$}{#1}\global\advance\notenumber by 1}
\def\abstract#1{\centerline{\null}
                \oneskip\oneskip
                \setbox\abstr=\vbox{\hsize 5.5truein{\par#1}}
                \centerline{SUMMARY} \oneskip \hbox to
                \hsize{\hfill\box\abstr\hfill}}
\def\new{{\rm\chaphead\the\eqnumber}\global\advance\eqnumber by 1}
\def\eqref#1{\advance\eqnumber by -#1 \chaphead\the\eqnumber
           \advance\eqnumber by #1 }
\def\last{\advance\eqnumber by -1 {\rm\chaphead\the\eqnumber}\advance
     \eqnumber by 1}
\def\eqnam#1{\xdef#1{\chaphead\the\eqnumber}}
\def\nfig{\chaphead\the\fignumber\global\advance\fignumber by 1}
\def\nfiga#1{\chaphead\the\fignumber{#1}\global\advance\fignumber by 1}
\def\rfig#1{\advance\fignumber by -#1 \chaphead\the\fignumber
            \advance\fignumber by #1}
\def\fignam#1{\xdef#1{\chaphead\the\fignumber}}


\def\part#1#2{{\partial#1\over\partial#2}}

\def\frac#1#2{{\textstyle{#1\over#2}}}

\linespacingstep=0 \multiplelines \normalbaselines
\ifnum\linespacingstep=0
  \footline{\hfil}
  \headline{\ifnum\pageno=1
    \hfil To appear in {\it The Astrophysical Journal}, March 20, 1996\/
  \else
    {\headfont Triaxial galaxies with cusps} \hfil \twelverm\folio \hfil
    Merritt \& Fridman 
  \fi}
\else
  \hsize=16cm
  \vsize=21cm
\fi

\newbox\topbox

\long\def\mtopbox#1{\setbox\topbox=\vbox{#1}
  \topinsert{\box\topbox}\endinsert}

\mtopbox{\ifnum\linespacingstep=0 \vskip 0.5cm \else \vskip 5cm \fi

\centerline{\uppercase{\titlefont Triaxial Galaxies with Cusps}} }
\ifnum\linespacingstep=0 \blankline \else \vskip 2cm \fi
\centerline{\authorfont David Merritt and Tema Fridman}
\vskip-3pt
\centerline{\affilfont Department of Physics and Astronomy, Rutgers
University, Piscataway, NJ 08855}

\ifnum\linespacingstep=0
  \blankline
  \centerline{Rutgers Astrophysics Preprint No 160}
\fi

\ifnum\linespacingstep=0 \blankline \else \vfill\eject \fi
{\narrower
\centerline{ABSTRACT}
We have constructed fully self-consistent models of triaxial galaxies with
central density cusps.
The triaxial generalizations of Dehnen's (1993) spherical mass
models are presented, which have
densities that vary as $r^{-\gamma}$ near the center and $r^{-4}$
at large radii.
We computed libraries of $\sim 7000$ orbits in each of
two triaxial models with
$\gamma=1$ (``weak cusp'') and $\gamma=2$ (``strong cusp'');
these two models have density profiles similar to those of the
``core'' and ``power-law'' galaxies observed by HST.
Both mass models have short-to-long axis ratios of 1:2 and are
maximally triaxial.
The major orbit families and their associated periodic orbits were
mapped as a function of energy.
A large fraction of the orbits in both model potentials are
stochastic, as evidenced by their non-zero Liapunov exponents.
We show that most of the stochastic orbits in the strong-cusp potential
diffuse relatively quickly through their allowed phase-space
volumes,
on time scales of $10^2 - 10^3$ dynamical times.
Stochastic orbits in the weak-cusp potential diffuse more slowly,
often retaining their box-like shapes for $10^3$ dynamical times
or longer.

Attempts to construct self-consistent solutions using just the regular
orbits failed for both mass models.
Quasi-equilibrium solutions that include the stochastic orbits exist for
both models; however, real galaxies constructed in this way
would evolve near the center due to the continued mixing of the
stochastic orbits.
We attempted to construct more nearly stationary models in which
stochastic phase space was uniformly populated at low energies.
These ``fully mixed'' solutions were found to exist only for the
weak-cusp potential; as much as $\sim 1/3$ of the
mass near the center of these models could be
placed on stochastic
orbits without destroying the self-consistency.
No significant fraction of the mass could be placed on
fully-mixed stochastic orbits in the strong-cusp model,
demonstrating that strong triaxiality can be inconsistent
with a high central density.

Our results suggest that chaos is a generic feature of the
motion in realistic triaxial potentials, but that the presence of
chaos is not
necessarily inconsistent with the existence of stationary triaxial
configurations.

\vskip 3pt

\ifnum\linespacingstep=0 \blankline \else \vskip 1cm \fi
 \par}

\vfill\eject

\sect{INTRODUCTION}

A central question in the study of elliptical galaxies is the extent
to which nature can construct equilibrium stellar systems that are not
axisymmetric.
Schwarzschild (1979, 1982) presented two numerical models of equilibrium
triaxial galaxies, with and without figure rotation, and Statler (1987)
showed that self-consistent equilibria exist for the fully integrable,
``perfect'' mass models presented by Kuzmin (1973) and by de Zeeuw
\& Lynden-Bell (1985).
Triaxial stellar systems also seem to form naturally in $N$-body
simulations (e.g. Aarseth \& Binney 1978; Wilkinson \& James 1982).

Real elliptical galaxies look rather different from the idealized
models of Schwarzschild (1979) and Statler (1987), which had large,
constant-density cores in which the orbital motion is essentially
harmonic.
Recent ground based (Moller, Stiavelli \& Zeilinger 1995) and space
telescope (Crane \etal~1993; Jaffe \etal~1994; Ferrarese \etal~1994;
Lauer \etal~1995)
observations demonstrate that early-type galaxies essentially never
have constant-density cores; the stellar surface
brightness always continues to rise into the smallest observable radius.
(Systematic deviations of elliptical galaxy core profiles from
the ``isothermal'' law were noted as early as 1985 by Kormendy from
ground-based observations.)
Ferrarese \etal~ (1994) and Lauer \etal~ (1995)
divide elliptical
galaxies into two classes based on their nuclear properties,
which Lauer \etal~ call ``core'' and ``power-law'' galaxies.
``Core'' galaxies exhibit an obvious break in the surface brightness
profile at some radius $R_b$; inward of this break, the profile turns
down to a shallow inner power law $\Sigma(R)\propto R^{-\alpha}$,
$\alpha\approx -0.1\pm 0.1$.
``Power-law'' galaxies show roughly a single power-law profile
throughout their inner regions with $\alpha\approx -0.8\pm 0.2$.
Power-law galaxies are of lower average luminosity
than core galaxies, but steep surface brightness profiles are seen
in galaxies with a very wide range of luminosities, from $M_V\approx
-15$ to $M_V\approx -22$ (Kormendy \etal~ 1995).

The division of elliptical galaxies into two groups based on
their surface brightness profiles is probably unnecessary once
one takes into account the effects of projection on the luminosity
density profiles.
Nonparametric deprojection of the Lauer {\it et al.} surface brightness
data (Merritt \& Fridman 1995) demonstrates that the so-called
``core'' galaxies, like the ``power-law'' galaxies, also exhibit
power-law cusps in their {\it spatial} densities.
These galaxies appear in projection to have ``cores''
only because the logarithmic slopes of their central luminosity density
profiles lie between $-1$ and $0$, and power-law cusps shallower
than $1/r$ do not appear as power laws when seen in projection
(e.g. Dehnen 1993, Fig. 1).
Lauer {\it et al.}'s ``power-law'' galaxies, by constrast, have
spatial density cusps with power-law indices that lie between
$-1$ and $-2$; cusps this steep retain their power-law character
even when projected against the outer layers of the galaxy.
Gebhardt {\it et al.} (1996) have recently shown that there is no
qualitative difference between the luminosity density profiles
of the two types of galaxy, and in fact both can be well fit by a single
family of parametric models, such as the family of Dehnen (1993) that
is used here (Kormendy, private communication).
Thus, power-law density cusps are a generic feature of
elliptical galaxies, and the fact that some ellipticals have
surface brightness profiles that flatten at small radii
appears to be an artifact of projection onto the plane of the sky.

This universal, power-law character of elliptical galaxy nuclei
is not wholly surprising in view of the fact that $N$-body simulations
of structure formation often produce systems with power-law
density profiles (e.g. Crone, Evrard \& Richstone 1994).

It has long been known that the addition of a central mass
concentration to an otherwise integrable triaxial potential can have
profound effects on at least one family of orbits, the boxes, which
fill a region that includes the center.
Gerhard and collaborators (Gerhard \& Binney 1985; Gerhard 1987)
showed that nuclear black holes or density cusps will subject stars on
box orbits to deflections that can destroy two of their three integrals of
motion.
The evolution of such an orbit can be described as a series of
near-random transitions from one box orbit to another;
after a long time, the orbit
would be expected to uniformly fill the phase-space region
corresponding to all trajectories, with a given energy,
that pass near the center.
Gerhard \& Binney (1985) pointed out that the time-averaged density
distribution of
such an orbit is likely to be rounder than that of a typical box
orbit, which suggests that stationary triaxial configurations may not
exist when the central mass concentration is too high.
None of these objections apply to axisymmetric models, for which all
the orbits conserve one component of the angular momentum and
therefore avoid the center.
Thus Gerhard \& Binney (1985) suggested that triaxial galaxies with
strong central mass concentrations would evolve in the
direction of axisymmetry, at least near their centers, as the box
orbits gradually lost their distinguishability.

One complication to this picture is the existence of stable
box-like orbits.
Periodic orbits are dense in any phase space, and periodic orbits that
avoid the center can remain stable even in the presence of a central
cusp or black hole.
Examples of periodic orbits that (like box orbits) begin on an
equipotential surface and pass close to the center are the 2:1
planar orbits, the ``bananas;'' the 3:2 ``fish;'' the 4:3
``pretzels;'' etc.
If they are stable, such periodic orbits can generate families
of regular orbits that might
contribute strongly to a self-consistent solution.
Schwarzschild and collaborators (Miralda-Escud\'e \& Schwarzschild
1989; Lees \& Schwarzschild 1992) investigated the existence and
stability of these ``boxlets'' in the principal planes of a variety of
non-integrable triaxial models, including the logarithmic potential
that corresponds to an inverse-square dependence of density on radius.
Based on the variation of boxlet shape with model flattening, these
authors concluded that strongly flattened, triaxial models might not
exist.
Pfenniger \& de Zeeuw (1989) reached a similar conclusion.

Kuijken (1993) constructed self-consistent, scale-free models of
triaxial disks, and Schwarzschild (1993), extending the work of
Richstone (1980, 1982, 1984) and Levison \& Richstone (1987),
found self-consistent solutions for a
set of three-dimensional scale-free ($\rho\propto r^{-2}$)
mass models with various axis ratios, designed to represent dark halos.
The orbital motion in Schwarzschild's models was
largely stochastic, somewhat more so than suggested by
earlier surface-of-section studies in the principal planes of
scale-free models (e.g. de Zeeuw \& Pfenniger 1988;
Miralda-Escud\'e \& Schwarzschild 1989).
(We demonstrate below that this enhanced stochasticity can be traced to
the vertical instability of the planar periodic orbits.
Motion in the principal planes turns out to be a poor guide
to the three-dimensional motion.)
Schwarzschild (1994) found that self-consistent solutions could not
always be constructed using only the regular orbits, especially for
flatter models.
However solutions that included the stochastic orbits could always be
found.
We will use the term ``quasi-equilibrium'' to describe solutions
like Schwarzschild's that contain distinguishable stochastic orbits.
In a quasi-equilibrium model, the continued mixing of the
stochastic orbits would produce a slow evolution of the model figure,
especially near the center.
Schwarzschild estimated that this evolution would not seriously
compromise the self-consistency of his models, at least over time scales
corresponding to $\sim 10^2$ orbital periods that are relevant to the
outer parts of galactic halos.
Orbital periods near the center of a galaxy with a cusp can
be much shorter than 1\% of a Hubble time, however, and the
evolution of
stochastic orbits would be expected to have more serious
consequences there.

Here we present the first, fully self-consistent, non-scale-free models of
triaxial galaxies with central density cusps.
Motivated by the space telescope observations described above,
we investigate two mass models with densities that increase as
$1/r$ (``weak-cusp'') and $1/r^2$ (``strong-cusp'') near the center (\S 2);
both have minor-to-major axis ratios of 1:2 and are maximally triaxial.
We find  (\S 3) that the phase space
corresponding to orbits that touch an equipotential surface, and that would be
classified as ``box'' orbits in an integrable potential,
is largely stochastic for both mass models.
We investigate the time scales for diffusion of these stochastic
orbits and show that they are relatively short, roughly $10^3$ dynamical
times, in the model with the $1/r^2$ cusp (\S 4).
We then attempt (\S 5) to construct self-consistent solutions that include the
stochastic orbits in various ways.
Solutions that exclude the stochastic orbits do not exist for either
mass model - the regular orbits are sufficiently rare, and limited
enough in their shapes, that they can not reproduce the model density
everywhere.
Quasi-equilibrium solutions, in which the stochastic orbits are
treated like the regular orbits, exist for both mass models; however,
a galaxy constructed in this way would evolve in shape,
especially near the center, as the stochastic orbits continued to
diffuse through their allowed phase-space regions.

We therefore attempted to construct models that exclude the
stochastic orbits at low energies, where the diffusion time
scales are short, or that populate the
stochastic orbits in an approximately time-independent way.
We found that these ``fully mixed'' solutions exist for the weak-cusp
model, in the sense that the stochastic orbits throughout the central
regions of the model could be replaced by a single ``orbit'' at
every energy
representing a  steady-state population of phase space
without destroying the self-consistency.
A galaxy constructed in this way would evolve only very slowly due to
the continued mixing of stochastic orbits at large energies.
Fully-mixed solutions could not be found for the model with the
$1/r^2$ cusp, however, demonstrating that strong central mass
concentrations are sometimes inconsistent with triaxiality.
The smaller variety of self-consistent solutions for the
strong-cusp model is probably due to the paucity of regular orbit
families in this potential; hence, more weight is placed on the
stochastic orbits.

All of our self-consistent solutions are dominated by stochastic
orbits.
Second in importance are the tube orbits, which
typically contribute about half of the total mass;
the ``boxlets,'' the regular remnants of the box orbits,
are dynamically almost insignificant.
Our results highlight the likely importance
of chaos in the phase-space structure of strongly triaxial stellar systems.
We show (\S 5) how Jeans's theorem can be generalized to include systems
containing both regular and stochastic orbits; thus,
chaos is not necessarily inconsistent with fully
stationary triaxial equilibria.
However our attempts at model-building demonstrate that
fully stationary equilibria do not exist for every choice of
triaxial mass model, including some that are quite similar in
appearance to real galaxies.
Furthermore, nature may not be inclined to populate the
stochastic orbits in just the right way to guarantee full
equilibrium, especially at large radii where mixing time
scales for stochastic orbits are long.
Thus, slow evolution may be a generic property of triaxial
stellar systems (\S 6).

\sect{Density, Potential, Forces}

The mass models considered in this study are the triaxial generalizations of
the spherical models first discussed in detail by Dehnen (1993), and more
recently by Carollo (1993) and Tremaine {\it et al.} (1994).
Our models have a mass density
$$\eqnam{\etam}
\rho(m) = {(3-\gamma) M\over 4\pi abc} m^{-\gamma} (1+m)^{-(4-\gamma)},
\ \ \ \ 0\le\gamma < 3 \eqno (\new a)
$$
with
$$
m^2={x^2\over a^2} + {y^2\over b^2} + {z^2\over c^2}, \ \ \ \ a\ge b\ge c\ge
0, \eqno (\last b)
$$
and $M$ the total mass.
The mass is stratified on ellipsoids with axis ratios $a:b:c$; the $x$ [$z$]
axis is the long [short] axis.
The parameter $\gamma$ determines the slope of the central density cusp.
For $\gamma=0$ the model has a finite-density core; for
$\gamma>0$ the
central density is infinite.
The strongest cusp that we will consider here has $\gamma=2$, i.e.
$\rho\propto m^{-2}$ at small radii.
At large radii, all models have $\rho\propto m^{-4}$.

The functional form (\etam), with a suitable choice for $\gamma$, has
been shown to be a good approximation to the luminosity densities of a
number of elliptical galaxies.
Jaffe (1983) advocated the choice $\gamma=2$, and Hernquist (1990)
suggested $\gamma=1$.
As discussed above, recent space telescope observations demonstrate
that elliptical galaxies and bulges exhibit a variety of cusp
strengths, $\gamma\ltsim 2$.
Few if any galaxies are observed to have $\gamma\approx 0$, i.e. a
constant-density core.

The gravitational potential of a body in which $\rho = \rho(m^2)$
may be written
$$
\Phi({\bf x}) = -\pi G abc \int_0^{\infty} {\left[\psi(\infty) -
\psi(m)\right]d\tau\over\sqrt{(\tau +a^2)(\tau+b^2)(\tau+c^2)}} \eqno(\new a)
$$
(Chandrasekhar 1969, Theorem 12), with
$$
\psi(m) = \int_0^{m^2} \rho(m'^2)dm'^2 \eqno(\last b)
$$
and
$$
m^2(\tau) = {x^2\over a^2+\tau} + {y^2\over b^2+\tau} + {z^2\over c^2+\tau} .
\eqno(\last c)
$$
We find, for $\gamma\ne 2$,
$$\eqnam{\potential}
\Phi({\bf x}) = -{GM\over 2(2-\gamma)}\int_0^{\infty}
{\left[1-(3-\gamma)\left({m\over
1+m}\right)^{2-\gamma}+(2-\gamma)\left({m\over 1+m}\right)^{3-\gamma}\right]
d\tau \over
\sqrt{(\tau +a^2)(\tau+b^2)(\tau+c^2)}}, \eqno (\new a)
$$
while for $\gamma=2$
$$
\Phi({\bf x}) = -{GM\over 2}\int_0^{\infty}{\left[\log\left({1+m\over
m}\right) - {1\over 1+m} \right]d\tau\over\sqrt{(\tau
+a^2)(\tau+b^2)(\tau+c^2)}}.
\eqno (\last b)
$$
In the spherical limit, these expressions reduce to those in Dehnen
(1993):
$$\eqalign{
\Phi(r) &= -{GM\over (2-\gamma)a}\left[1-{(r/a)^{2-\gamma}\over
(1+r/a)^{2-\gamma}}\right], \ \ \gamma\ne 2,\cr
&= -{GM\over a}\log\left(1+{a\over r}\right), \ \ \ \ \ \ \ \ \ \ \ \ \ \ \
\ \ \ \gamma=2.\cr} \eqno(\new)
$$
The central value of the potential is divergent for $\gamma=2$.

The gravitational forces are
$$\eqnam{\forces}
-{\partial\Phi\over\partial x_i} = -(3-\gamma) G \left({x_i\over
2}\right) \int_0^{\infty}
{m^{-\gamma}\ d\tau \over (a_i^2+\tau)(1+m)^{4-\gamma}
\sqrt{(\tau+a^2)(\tau+b^2)(\tau+c^2)}} , \ \ \ \ \  i=1, 2, 3, \eqno (\new)
$$
with $a_1\equiv a$, $a_2\equiv b$, $a_3\equiv c$.
In the spherical case, we have
$$\eqnam{\forceeta}
-{\partial\Phi\over\partial r}=-{GM\over a^2}\left({r\over a}\right)^
{1-\gamma}\left(1+{r\over
a}\right)^{\gamma-3}. \eqno (\new)
$$

Below we construct self-consistent models based on just two choices
for the parameters $a,b,c$ and $\gamma$.
Both mass models have the same set of axis ratios:
$$
{c\over a} = {1\over 2}, \ \ \ T \equiv {a^2-b^2\over a^2-c^2} = {1\over
2},
\eqno (\new)
$$
i.e. both are ``maximally triaxial'' ellipsoids with major-to-minor
axis ratios of 2:1.
Model 1, which we will call the ``weak cusp'' model, has $\gamma=1$,
while model 2, the ``strong cusp'' model, has $\gamma=2$.
According to equation (\forceeta), the central force diverges as $r^{-1}$
in the strong cusp model, while in Model 1 the central force is
finite but nonzero.

Henceforth we adopt units in which the total mass $M$, the $x$-axis scale
length $a$, and the gravitational constant $G$ are unity.

The expressions (\potential) and (\forces) are improper integrals and not
very suitable for numerical computation.
The potential integrals were rewritten using the substitution
$s=(1+\tau)^{-1/2}$.
For $\gamma\ne 2$ this yields the proper integral
$$
\Phi = -{1\over 2-\gamma}\int_0^1{\left[1-(3-\gamma)\left({m\over
1+m}\right)^{2-\gamma} + (2-\gamma)\left({m\over 1+m}\right)^{3-\gamma}\right]
ds\over \sqrt{\left[1+(b^2-1)s^2\right] \left[1+(c^2-1)s^2\right]}}, \eqno
(\new a)
$$
with
$$
m^2(s)=s^2\left[x^2 + {y^2\over 1+(b^2-1)s^2} + {z^2\over
1+(c^2-1)s^2}\right]. \eqno (\last b)
$$
For $\gamma=2$ we have
$$\eqalignno{
\Phi& = -\int_0^1 {\left[\log\left({(1+m)s\over m}\right) - {1\over
1+m}\right] ds \over \sqrt{\left[ 1+(b^2-1)s^2\right]
\left[1+(c^2-1)s^2)\right]}} + C, &(\new a)\cr
C &= \int_0^1 {\log t\ dt\over\sqrt{\left[1+(b^2-1)t^2\right]
\left[1+(c^2-1)t^2\right] }}. &(\last b)\cr}
$$

The substitution $s=a_i(a_i^2+\tau)^{-1/2}$ gives for the components of the
force
$$
F_i=-{\partial\Phi\over\partial x_i} = -(3-\gamma) {x_i\over a_i} \int_0^1
{s^2 ds \over
m^{\gamma}(1+m)^{4-\gamma}\sqrt{\left(a_i^2+A_1s^2\right)
\left(a_i^2+A_2s^2\right)}}, \eqno (\new a)
$$
with
$$
m^2(s) = s^2\left( {x^2\over a_i^2+C_1s^2} + {y^2\over a_i^2+C_2s^2} +
{z^2\over
a_i^2+C_3s^2}\right). \eqno (\last b)
$$
The constants are
$$\eqalign{
i=1:\ &A_1=b^2-1\ \ \ A_2=c^2-1\ \ \ \ C_1=0\ \ \ \ \ \ \ \ \ C_2=b^2-1 \
\ \ C_3=c^2-1\cr
i=2:\ &A_1=c^2-b^2\ \ A_2=1-b^2\ \ \ \ C_1=1-b^2\ \ \ C_2=0\ \ \ \ \ \ \
\ \ C_3=c^2-b^2\cr
i=3:\ &A_1=1-c^2\ \ \ A_2=b^2-c^2\ \ \ C_1=1-c^2\ \ \ C_2=b^2-c^2 \
\ C_3=0.
\cr}
\eqno (\last c)
$$
The transformed integrals were evaluated numerically using the NAG routine
D01AHF.

The derivatives of the forces were needed when computing the Liapunov
exponents, as described below (\S 3).
These are:
$$
{\partial^2\Phi\over\partial x_i^2} = -{F_i\over x_i} - {3-\gamma\over
a_i} \left({x_i\over a_i}\right)^2 \int_0^1 {(\gamma+4m_i)s^4 ds\over
m_i^{\gamma+2} (1+m_i)^{5-\gamma}\sqrt{(a_i^2+A_1s^2)(a_i^2+A_2s^2)}},
\eqno (\new a)
$$
$$
{\partial^2\Phi\over\partial x_i\partial x_j} = - {(3-\gamma)x_i x_j\over
a_i} \int_0^1 {(\gamma+4m_i)s^4 ds\over
m_i^{\gamma+2} (1+m_i)^{5-\gamma}[a_i^2+(a_j^2-a_i^2)s^2]^{3/2}
\sqrt{(a_i^2+(a_k^2-a_i^2)s^2}},
\eqno (\new b)
$$

Because we forced our models to have equidensity surfaces that are
precisely ellipsoidal, our expressions for the forces and the potential
could not be reduced below one-dimensional integrals and were
therefore expensive to compute.
We justify this expense on the grounds that real elliptical galaxies
have isophotes that are also very nearly elliptical.
Much modelling in the past has been based on simpler expressions for
the forces that correspond to strongly dimpled
mass distributions.

In our adopted units, one unit of time corresponds to:
$$
1.49\times 10^6 {\rm yr} \left({M\over 10^{11}M_{\odot}}\right)^{-1/2}
\left({a\over 1{\rm kpc}}\right)^{3/2}. \eqno (\new)
$$
We will often present elapsed times in units of an energy-dependent
``dynamical time'' or ``orbital time'' $T_D$, defined as the period
of the (nearly circular) 1:1 resonant orbit in the $x-y$ plane.
Table 1 gives the period of this orbit in model units as a function of
energy for the two adopted mass models.
The energy values in this table are equal to those used for the
construction of the orbit library, as described below.

\sect{Integration of Orbits and Computation of Liapunov Exponents}

Because of the extreme inhomogeneity of these models, the numerical
algorithm for integrating the orbits must be extremely accurate and
flexible.
We used the 7/8 order
Runge-Kutta algorithm described by Fehlberg (1968), which incorporates a
variable time step in order to maintain a specified
accuracy from one integration step to the next.
A Fortran version of this routine, RK78, was kindly
made available by Dr. St\'ephane Udry.
The accuracy parameter TOL in all the integrations described below
was chosen to be $10^{-8}$.
We found that a typical orbit conserved energy to a few
parts in $10^9$ over 100 dynamical times with this choice of TOL -
a very high level of accuracy.

The potentials considered here are very different from the fully
integrable triaxial potentials discussed by Kuzmin (1973),
de Zeeuw (1985) and others.
We therefore expect that many of the trajectories will conserve only
one or two integrals of the motion, rather than the three integrals
that characterize fully regular motion.
These ``stochastic'' orbits behave qualitatively differently in
many respects from regular orbits and must be treated separately in
the construction of self-consistent models, as discussed in
greater detail below.

A standard way of detecting and quantifying stochasticity
is through computation of the Liapunov characteristic exponents (e.g.
Lichtenberg \& Lieberman 1992, p. 296).
The Liapunov exponents of a trajectory are the mean exponential
rates of divergence of trajectories surrounding it.
Consider a trajectory in six-dimensional phase space and a nearby
trajectory with initial conditions ${\bf x_0}$ and ${\bf x_0+\Delta
x_0}$, respectively.
These evolve with time yielding a difference vector ${\bf w}
({\bf x_0},t)$ with length $d({\bf x_0},t)$.
The mean exponential rate of divergence of two initially close
trajectories is
$$\eqnam{\liapdef}
\sigma({\bf x_0},\Delta{\bf x}) = \lim_{t\rightarrow \infty} \left({1\over
t}\right) \ln {d({\bf x_0},t)\over d({\bf x_0},0)}. \eqno (\new)
$$
It can be shown that there is a six-dimensional basis $\{\hat e_i\}$
of ${\bf w}$ such that for any ${\bf w}$, $\sigma$ takes on one of the
six values
$$
\sigma_i({\bf x_0}) = \sigma({\bf x_0},\hat e_i), \eqno (\new)
$$
which are the Liapunov exponents.
These can be ordered by size,
$$
\sigma_1\ge\sigma_2\ge...\ge\sigma_6.
$$
In the special case of Hamiltonian flow considered here, phase space
volume is a conserved quantity and so only three of the Liapunov
exponents are independent.
It may be shown that
$$
\sigma_i=-\sigma_{7-i}, \eqno (\new)
$$
so that we may henceforth restrict our attention to the three positive
exponents.
Furthermore, for Hamiltonian flow, one of these three exponents
will be exactly zero.

Each additional isolating integral aside from the energy causes one
more of the Liapunov exponents to be zero.
Regular orbits have three isolating integrals and their
Liapunov exponents are all zero; the trajectories around such orbits
diverge, at best, only linearly.
An orbit can be classified as stochastic if the greatest Liapunov
exponent is non-zero.
A stochastic trajectory that respects only one integral of motion, the
energy, will have two non-zero Liapunov exponents; a stochastic orbit that
respects two integrals (if such exist) will have only one
non-zero Liapunov exponent.

Liapunov characteristic exponents are defined as limiting values over an
infinite time interval.
Numerical approximations, computed over a finite time interval,
are sometimes called ``Liapunov characteristic indicators'' (e.g.
Heggie 1991).
We would expect all numerically-computed Liapunov exponents to
remain nonzero after a finite integration time.

We computed approximations to the six Liapunov exponents for each of
the orbits in our library using the Gram-Schmidt orthogonalization
technique described by Benettin \etal~ (1980).
A Fortran routine developed by the Geneva Observatory group
for carrying out the Benettin \etal~ algorithm, called LIAMAG, was
kindly made available by Dr. St\'ephane Udry.
The technique requires the integration of six perturbation orbits in
addition to the orbit under consideration (Udry \& Pfenniger
1988).
The evolution of the perturbed orbits is determined by the second
derivatives of the potential with respect to position; these
expressions are given in \S 2.
The time required to integrate an orbit including the six
perturbation orbits for 100
orbital times and compute the Liapunov exponents was typically
about 2.5 minutes on a DEC Alpha 3000/700 workstation.
Computation of the full set of 6840 orbits for one model
required about 250 hours.

{\epsfxsize 5.in \epsfbox{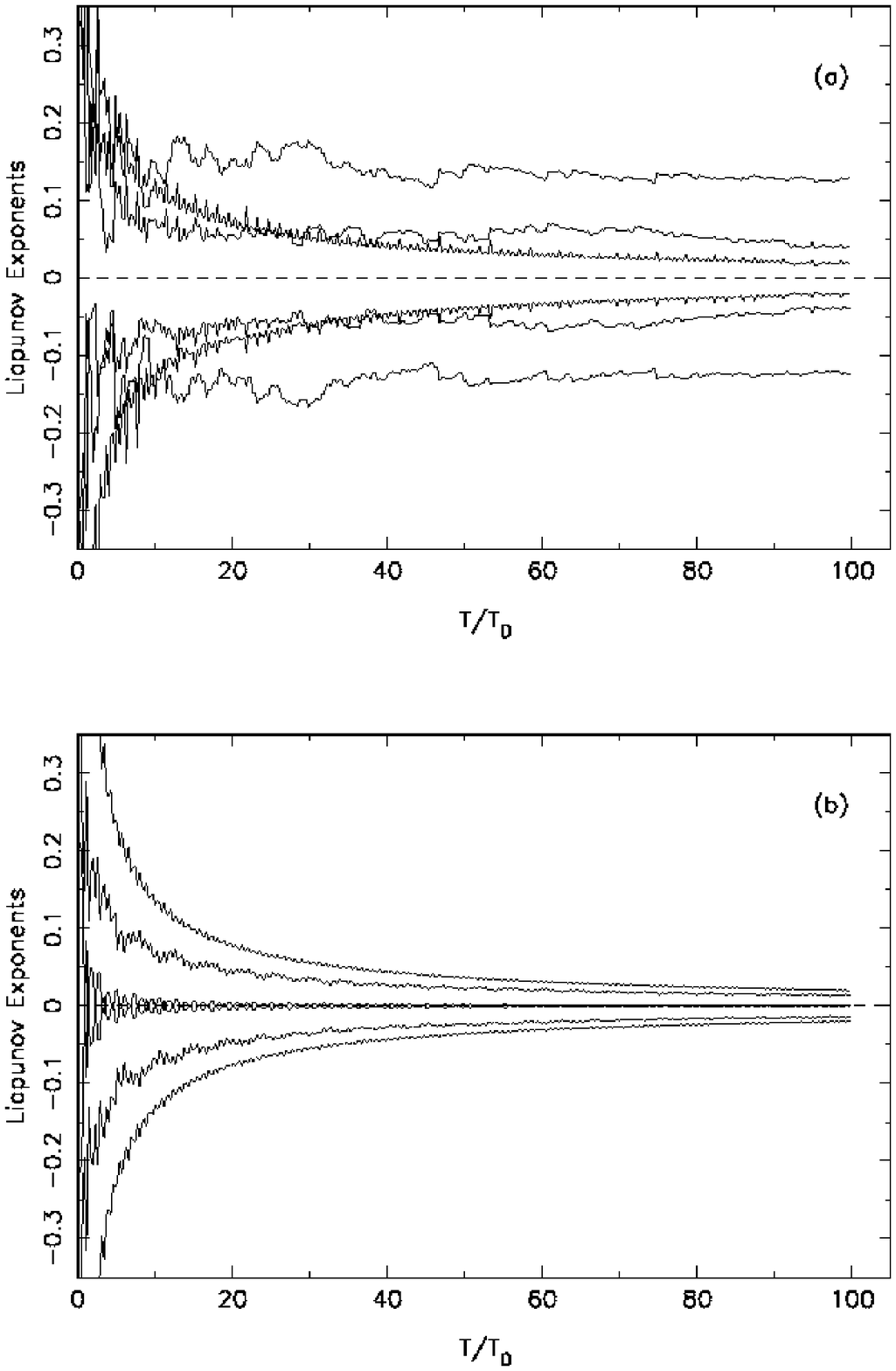}
{\normalbaselineskip=8pt\normalbaselines\noindent
Figure 1. Time dependence of the numerically-computed Liapunov
exponents for two orbits in the strong-cusp potential.
(a) Stochastic orbit; (b) regular orbit.
\par}

\bigskip

Figure 1 shows the numerically-computed Liapunov exponents as a function
of time for two orbits, one regular and one stochastic, in the strong-cusp
potential.
For these two orbits, the time dependence of the exponents is
reasonably clear after 100 dynamical times and
there is little ambiguity in classifying them as regular or
stochastic.

However for many orbits the situation is less clear.
The Liapunov exponents are defined as limiting values over an infinite
integration time (equation \liapdef).
A stochastic orbit that begins in a region of phase space that is
dominated by regular orbits will often remain ``trapped'' by the
surrounding tori for many oscillations, before ``breaking
out'' into the larger stochastic web.
Contopoulos \& Barbanis (1989) have emphasized that in weakly
stochastic regions of phase space, trajectories may need to be
integrated for as many as $10^5$ dynamical times or more in order
for the Liapunov exponents to be accurately determined.

Given these difficulties, we adopted the following procedure for
distinguishing regular from stochastic orbits.
A set of orbits of a single energy was integrated for 100 dynamical times,
with initial conditions chosen in a manner to be described below
(\S4).
A histogram was constructed of the positive Liapunov exponents of these
orbits: either the largest exponent $\sigma_1$; or the sum of the two largest
exponents $\sigma_1+\sigma_2$; or the sum of all three
positive exponents, sometimes called the ``Kolmogorov entropy,'' or $h_K$.
(Our finite-time approximations to $\sigma_3$ tended to be larger
for stochastic orbits than for regular ones; after very long
integration times, of course, $\sigma_3=0$ even for stochastic
orbits.)
The histogram based on $h_K$ was found to be the most effective in
distinguishing between regular and stochastic orbits.
If the Liapunov exponents were computed accurately, we would expect
these histograms to have just two strong peaks: one peak at zero,
corresponding
to the regular orbits; and one peak at some non-zero value, corresponding
to the stochastic orbits.
(This expectation is based on the hypothesis - not yet proved in the
general case - that all the stochastic
trajectories at a given energy in three-dimensional potentials are
interconnected via
the ``Arnold web.'')
A spread in values would indicate that the orbits have not been
integrated long enough to reach accurate limiting values
for the exponents.

Figure 2 shows frequency functions of the Kolmogorov entropy,
$h_k=\sum_{i=1,3}\sigma_i$, for two sets of 192
orbits computed for 100 dynamical times in the two adopted potentials.
(The frequency functions were computed using an adaptive kernel
algorithm that yields smooth, continuous curves.)
Each orbit began from a point on a single equipotential surface (shell
15 in the nomenclature defined below, \S 4).
The frequency function for orbits in the strong-cusp potential shows a clear
peak at small Liapunov numbers, corresponding to the regular orbits,
and a broader peak corresponding to the stochastic orbits.
In the weak-cusp potential, there is less of a clear distinction
between the two categories of orbits.

Clearly our integration time of 100 oscillations is barely sufficient to
separate regular from stochastic orbits in the weak-cusp potential,
and even in the strong-cusp case, the stochastic orbits show a wide
range of Liapunov numbers rather than the single sharp peak that we
would expect after an infinite integration time.
We therefore explored a number of alternative schemes for more reliably
distinguishing between the two sorts of orbits.

{\epsfxsize 5.in \epsfbox{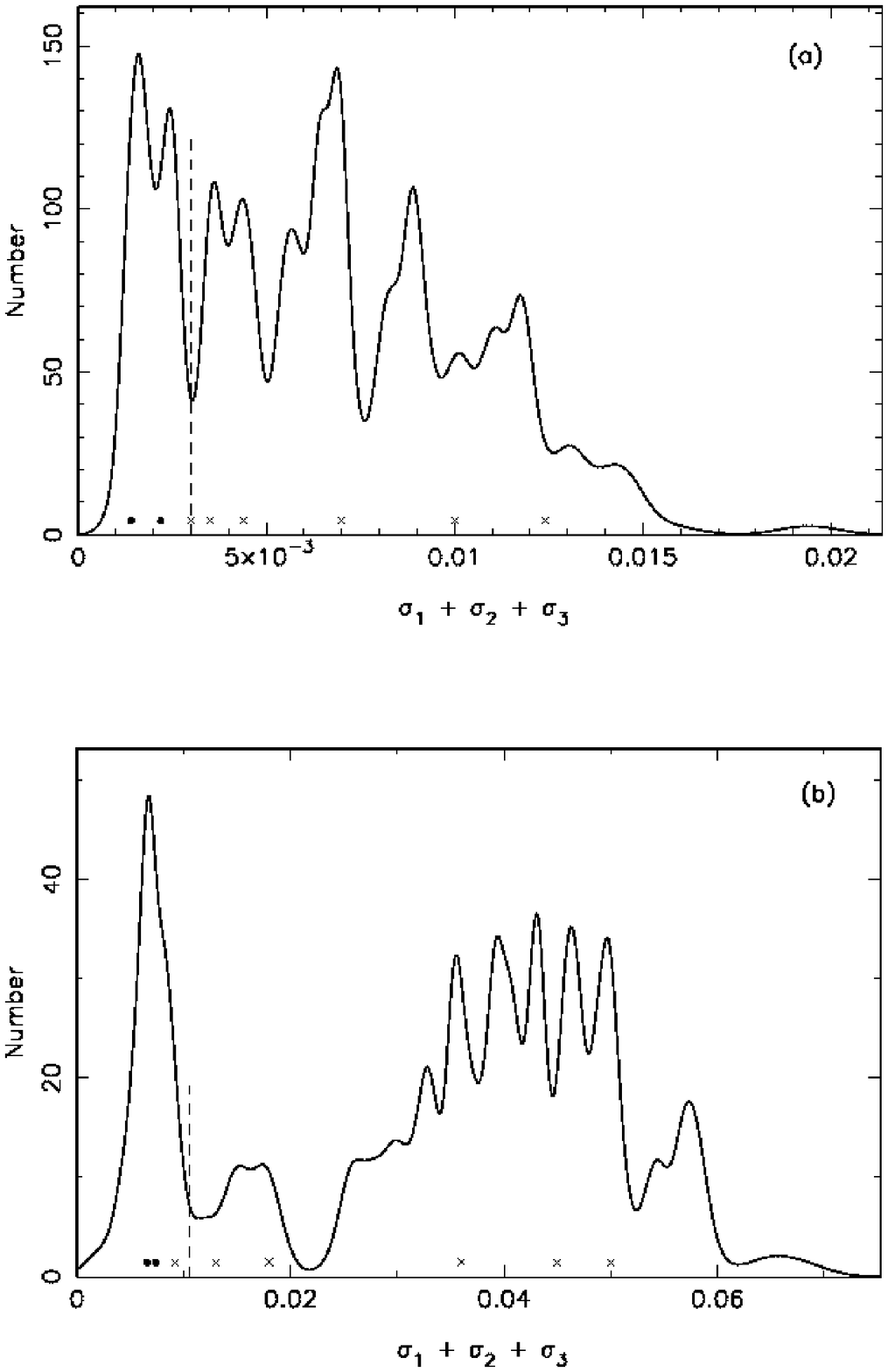}
{\normalbaselineskip=8pt\normalbaselines\noindent
Figure 2. Frequency function of the Kolmogorov entropy, $h_K=\sum_{i=1}^3
\sigma_i$, for 192 orbits from shell 15 of the stationary start
space.
(a) Weak-cusp model; (b) Strong-cusp model.
Dashed vertical line is the estimate of the critical $h_K$
separating regular (left) from stochastic (right) orbits.
$\bullet$ and $\times$ represent orbits that were integrated for
$10^5$ dynamical times; $\bullet =$ regular, $\times =$
stochastic.
\par}

\bigskip

We began by integrating a selected set of orbits at several energies for much
longer times, up to $10^5$ orbits, and observing their behavior using
the ``ergodicity index'' $\Delta$ defined in $\S4.2$.
The Liapunov exponents were not computed in order to save time.
These experiments showed that the Liapunov exponents computed after
only 100 orbits were a generally accurate indicator of stochasticity,
in the sense that a critical value of $h_K(t=100 T_D)$
could always be found that
more-or-less cleanly separated regular from stochastic orbits.
Although an occasional orbit with a very low $h_K$ was found to
eventually exhibit stochasticity, by and large the regular and
stochastic orbits lay on separate sides of the histogram.
Furthermore, the critical value of the Kolmogorov entropy
dividing regular from stochastic trajectories was almost always found
to lie close to a local minimum in the histogram, typically the first
minimum to the right of the peak containing the regular orbits.
A plot of critical $h_K$ versus energy showed a reassuringly smooth
and monotonic dependence.

Interestingly, the maximum Liapunov exponent $\sigma_1$,
or the sum of the two
largest exponents $\sigma_1+\sigma_2$, were found to be much less
useful for distinguishing between regular and stochastic trajectories
in this way.
Extra, and useful, information about the stochasticity is
apparently contained within the second and third exponents.

(One additional justification for computing the full set of Liapunov exponents
might be to determine whether the stochastic orbits respect just a
single isolating integral - the energy - or whether some orbits
respect two (not three, or one) isolating integrals.
In the latter case, we would expect to see just one, non-zero Liapunov
exponent.
Unfortunately the limited time over which we integrated our orbits did
not allow us to reach any very definite conclusions on this question.
However most stochastic orbits seemed to have two non-zero exponents
after 100 orbital times, implying the existence of no isolating
integrals aside from the energy.)

We also looked at a number of more qualitative indicators of
stochasticity, including plots of the configuration-space trajectory,
the dependence on time of the $\sigma_i$, etc.
We were struck by how well these different indicators agreed with
one another; for instance, a stochastic orbit could almost always be
identified as such based on an inspection of its configuration-space
trajectory over only a few tens of dynamical times.

Based on these experiments, we are confident that the majority of our
orbits have been correctly classified as regular or stochastic, in
spite of the rather short integration times.
There are undoubtedly some stochastic orbits that we have misclassified as
regular, due to ``trapping'' by nearby tori; some of
these stochastic orbits would require an exponentially long time
to exhibit their instability.
But there are probably few if any regular orbits that we have
misclassified as stochastic.
Thus, we have probably underestimated slightly the fraction of stochastic
orbits in our models.

The orbital times in the outer parts of real galaxies are
long enough that many stochastic orbits would behave essentially like
regular orbits over astronomical time scales.
One might therefore argue that the behavior of a
stochastic orbit over time scales much longer than 100 oscillations is
of little practical importance to the construction of
self-consistent models.
However dynamical times become much shorter near the centers of
galaxies, particularly galaxies that have central density cusps.
For this reason it is important to make a clear distinction between
regular and stochastic orbits even if the two classes of orbits
behave similarly on time scales of a few hundred oscillations.
We return to the question of diffusion time scales below.

\sect{Orbit Families}

For each of the two mass models considered here, 6840 orbits were
integrated for 100 dynamical times.
We followed Schwarzschild (1993) in assigning initial conditions from
one of two sets of starting points: either on an equipotential with
zero initial velocity (``stationary''); or in the $x-z$ plane with
$v_x=v_z=0$ (``$X-Z$'').
Orbits started in the $x-z$ plane are mostly tube orbits, i.e. they
avoid a region near the center of the model.
Orbits started on an equipotential surface tend to approach the
origin after sufficient time, except for those orbits that lie near
to a ``centrophobic'' periodic orbit.
Thus, orbits begun on an equipotential are either stochastic,
or else regular ``boxlets,'' i.e. associated with a stable periodic
orbit.
As Schwarzschild (1993) argues, these two initial condition spaces
probably include most - though strictly speaking not all - of the
orbits in the full phase space of a nonrotating triaxial model.

Orbits in both ``start spaces'' were assigned one of a set of 20
energies, defined as the values of the potential on the $x$-axis of a set of
ellipsoidal shells - with the same axis ratios as the density - that
divide the model into 21 sections of equal mass.
These energies are given in Table 1.
Thus shell 1 encloses 1/21 of the total mass, shell 2 encloses 2/21,
etc.
Shell 21 lies at infinity.
The $X-Z$ start space was defined as follows.
Let $\theta=\tan^{-1} x/z$ and pick 10 evenly spaced
values, from $\theta=2.25^{o}$ to $\theta=87.75^{o}$.
Compute for each $\theta$ the zero-velocity limit $\Phi(x,0,z)=E$
corresponding to the specified energy.
Define a circle in the $x-z$ plane whose radius is the minimum of
the amplitudes of the $x-y$ or $y-z$ 1:1 periodic orbits at that energy.
Then pick 15 equally spaced values at each $\theta$ at the centers of
15 equal intervals between this circle and the equipotential surface.
This scheme yields 150 orbits per shell and minimizes the duplication of
orbits that would result from selecting starting points throughout the
entire $x-z$ quadrant.

The stationary start space grid was defined as in Schwarzschild
(1993), except that each of the three sectors on an equipotential
octant contained only 64 initial points, for a total of 192
orbits per shell.

Orbits were integrated over a time interval equal to 100 dynamical
times as given in Table 1.
For each orbit, a detailed plot of the following quantities was made:

1. Projection of the orbit onto the $x=0$, $y=0$ and $z=0$ planes.

2. Dependence of the $x$ and $z$ components of the angular momentum on
time.

3. Liapunov exponents (i.e. their numerical approximations)
as a function of time.

4. Dependence of the energy conservation on time.

\noindent Orbits were then assigned to various categories according to the
following scheme.
First, the orbit was classified as regular or stochastic.
Here we used the procedure outlined above, which required plotting the
histogram of Kolmogorov entropies of all the orbits at a given energy
and estimating the critical value separating regular from stochastic
orbits.

Second, the regular orbits were assigned to one of a set of orbit families.
Clearly the possible number of such families is infinite, since
periodic orbits of ever higher order are dense in the phase space, and
any stable periodic orbit can act as the parent of nearby quasi-periodic
orbits.
However in practice only a subset of these periodic orbits - though
sometimes of surprisingly high order - were found to be important.

\subsect{Regular Orbits}

Our basic set of families included three types: long-axis tubes
(L), short-axis tubes (S), and boxes (B).
Long-axis tubes have a definite component of angular momentum about
the $x$-axis; short axis tubes have a definite $L_z$; and boxes have
no obvious, nonzero, time-averaged angular momentum components.

When possible, more detailed classifications were made.
Long-axis tubes were divided into inner (I) and outer (O)
families, following the terminology used by Kuzmin (1973) and de Zeeuw
(1985) for tube orbits in fully integrable potentials.
Not surprisingly, the presence of a central cusp has little effect on
orbits which avoid the center and we found that most long-axis tube
orbits fall clearly into the I or O families as defined by those
authors.

In the stationary start space, box orbits could often be
further grouped into families associated with a stable periodic orbit.
The most important of these periodic orbits were identified, and
their stability checked, using a Fortran routine
written by the Geneva group and kindly made available by S. Udry.
Tables 2-5 describe the major families of periodic orbits as a
function of energy in the two models.
Only the stable periodic orbits are included, with the exception of
some unstable low-order resonances in the principal planes, the 2:1
``bananas'' and the 3:2 ``fish''.
Entries for these low-order unstable orbit families are included at
shell 1 only.
An entry of `v' in the last column denotes instability out of a
principal plane, while `u' denotes instability in a more general
direction.

Figures 3-6 illustrate the two start spaces for each model.
The $X-Z$ start spaces are similar to those seen in integrable or
near-integrable models (e.g. Schwarzschild 1993); most orbits are
regular tubes, and fall into one of the three tube families
defined above.
The stationary start spaces for the weak-cusp model are complex,
containing large numbers of orbit families each of which
dominates a small part of the space.
At large energies the number of important resonances decreases,
leaving finally only the 2:1 $x-z$ banana family.
In the strong-cusp model, this family is dominant at all
energies; the variety of important resonances is
smaller than in the weak-cusp model.

The weak-cusp potential is closer than the strong-cusp potential
to a fully integrable model, in the sense of having a more
nearly constant-density core.
In a fully integrable potential, there is only one family of orbits (the
boxes) at each energy in stationary start space; the periodic
orbits, although they fill the start space densely, generate no
new families.
Thus we should not be surprised to see a larger number of
periodic orbit families in the weak-cusp model, each of
which dominates a smaller part of the start space.

{\epsfxsize 5.5in \epsfbox{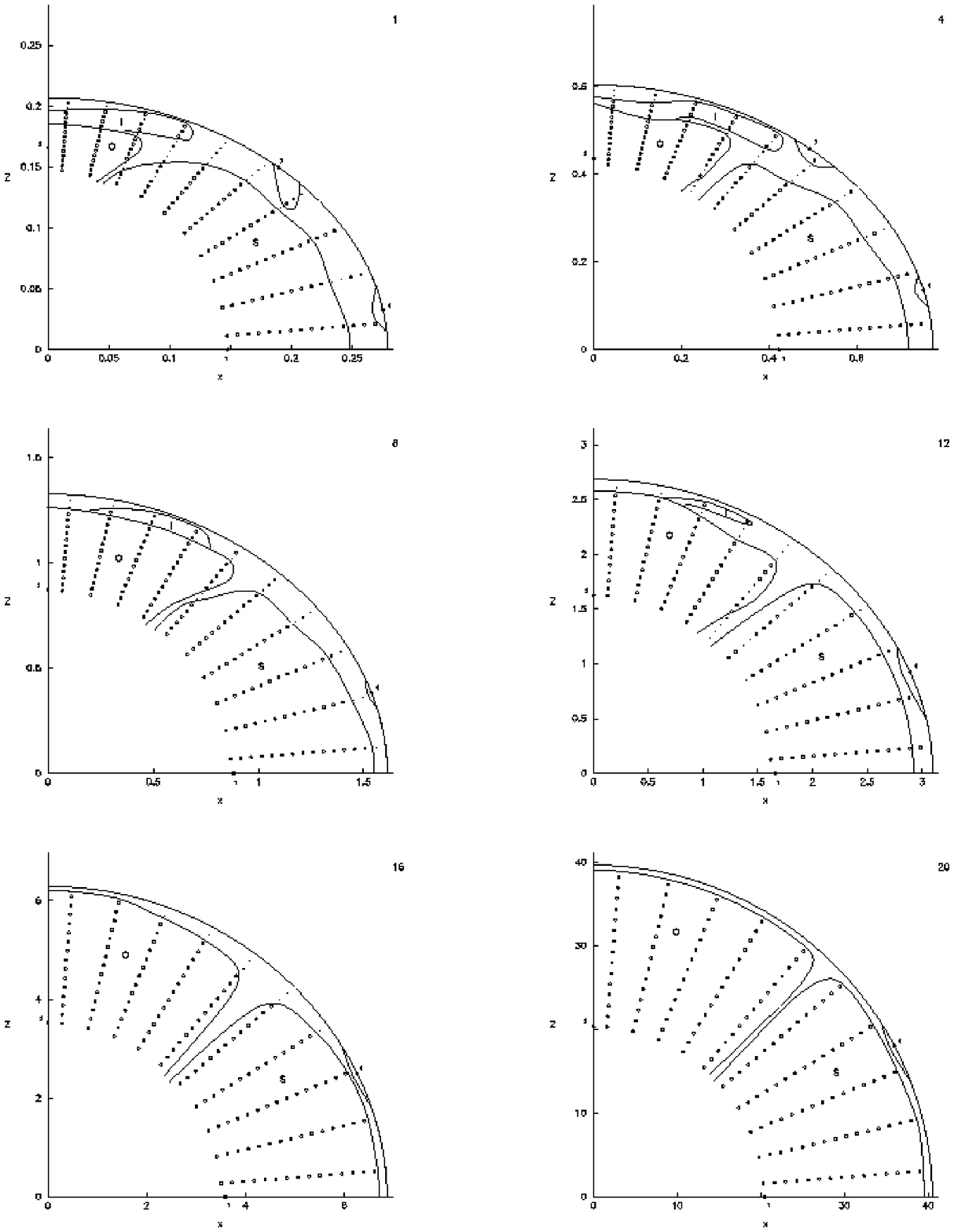}
{\normalbaselineskip=8pt\normalbaselines\noindent
Figure 3. The $X-Z$ start space at six energies for the weak-cusp model.
The numbers in the upper right of each frame denote the shell.
The open circles designate the regular orbits, and the small dots
the stochastic orbits, in a regular grid of
150 orbits.
The symbols S, O and I stand for the three major tube families.
The numbered heavy dots represent stable closed orbits for which
the resonances are listed in Table 2.
\par}
\vfill\eject

{\epsfxsize 5.5in \epsfbox{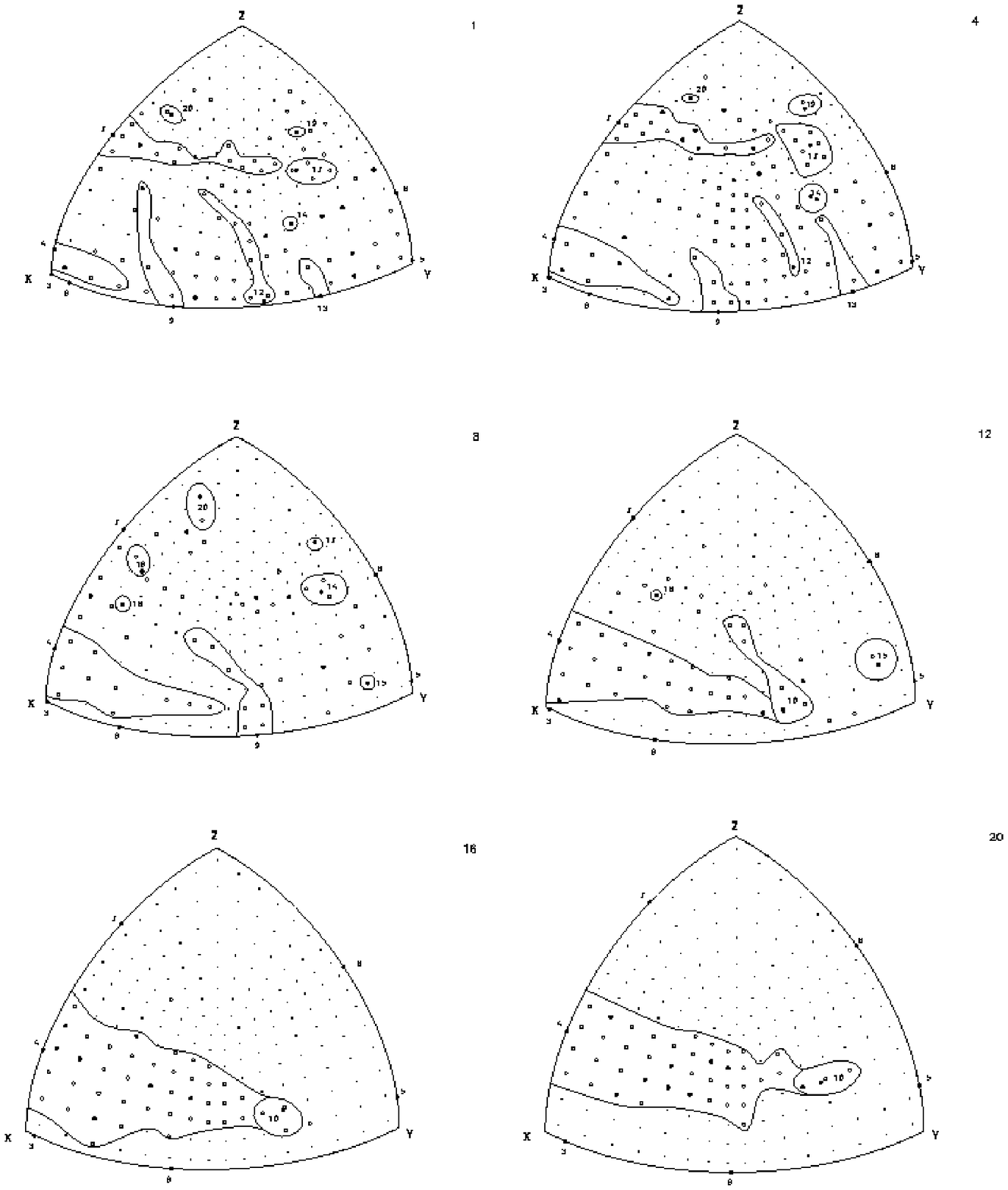}
{\normalbaselineskip=8pt\normalbaselines\noindent
Figure 4. The stationary start space at six energies for the
weak-cusp model.
The numbers in the upper right of each frame denote the shell.
The open circles designate the regular orbits, and the small dots
the stochastic orbits, in a regular grid of
192 orbits.
The numbered heavy dots represent stable closed orbits for which
the resonances are listed in Table 3.
\par}

\vfill\eject

{\epsfxsize 5.5in \epsfbox{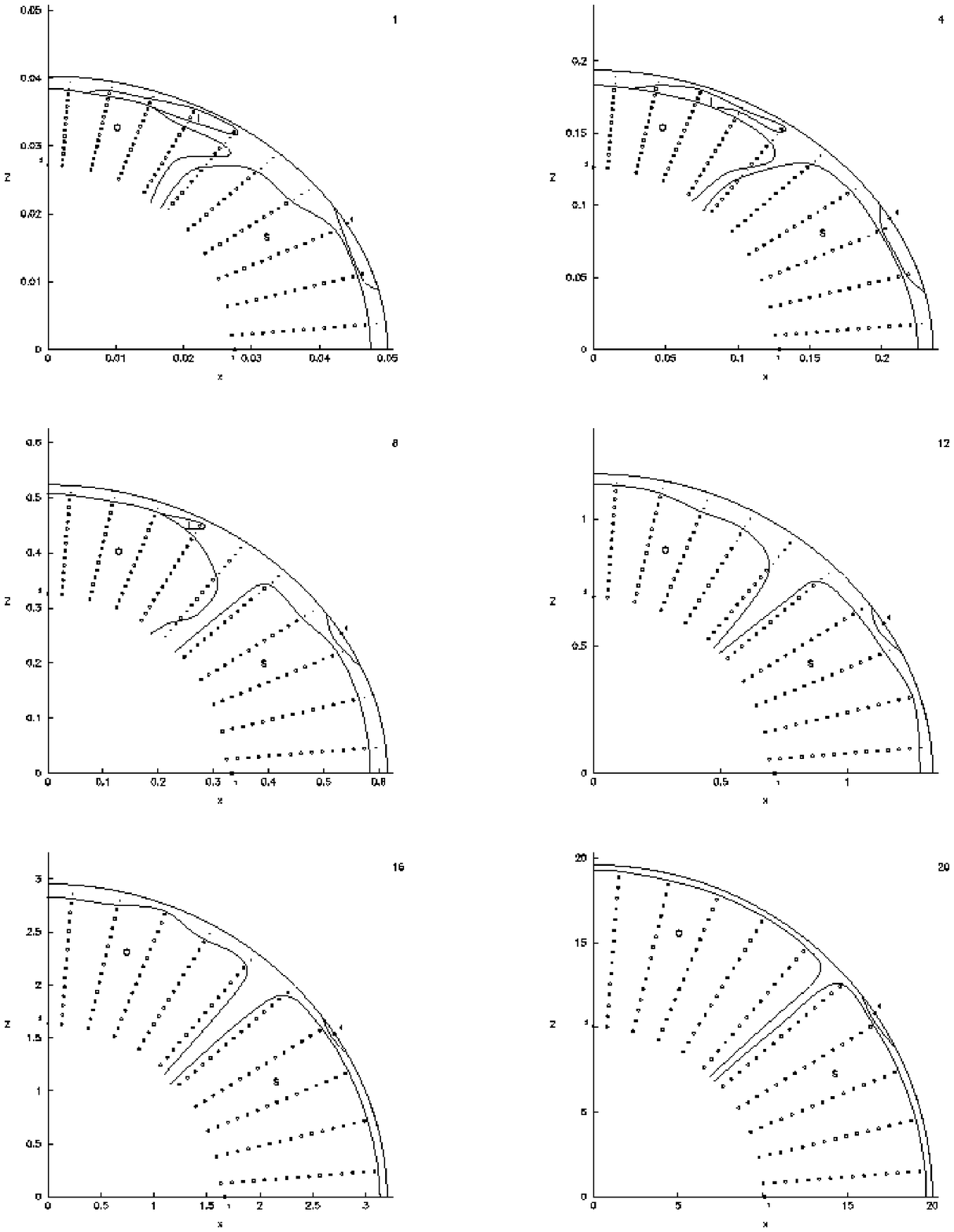}
{\normalbaselineskip=8pt\normalbaselines\noindent
Figure 5. Like Figure 3, for the strong-cusp model.
Resonant orbits are listed in Table 4.
\par}

\vfill\eject

{\epsfxsize 5.5in \epsfbox{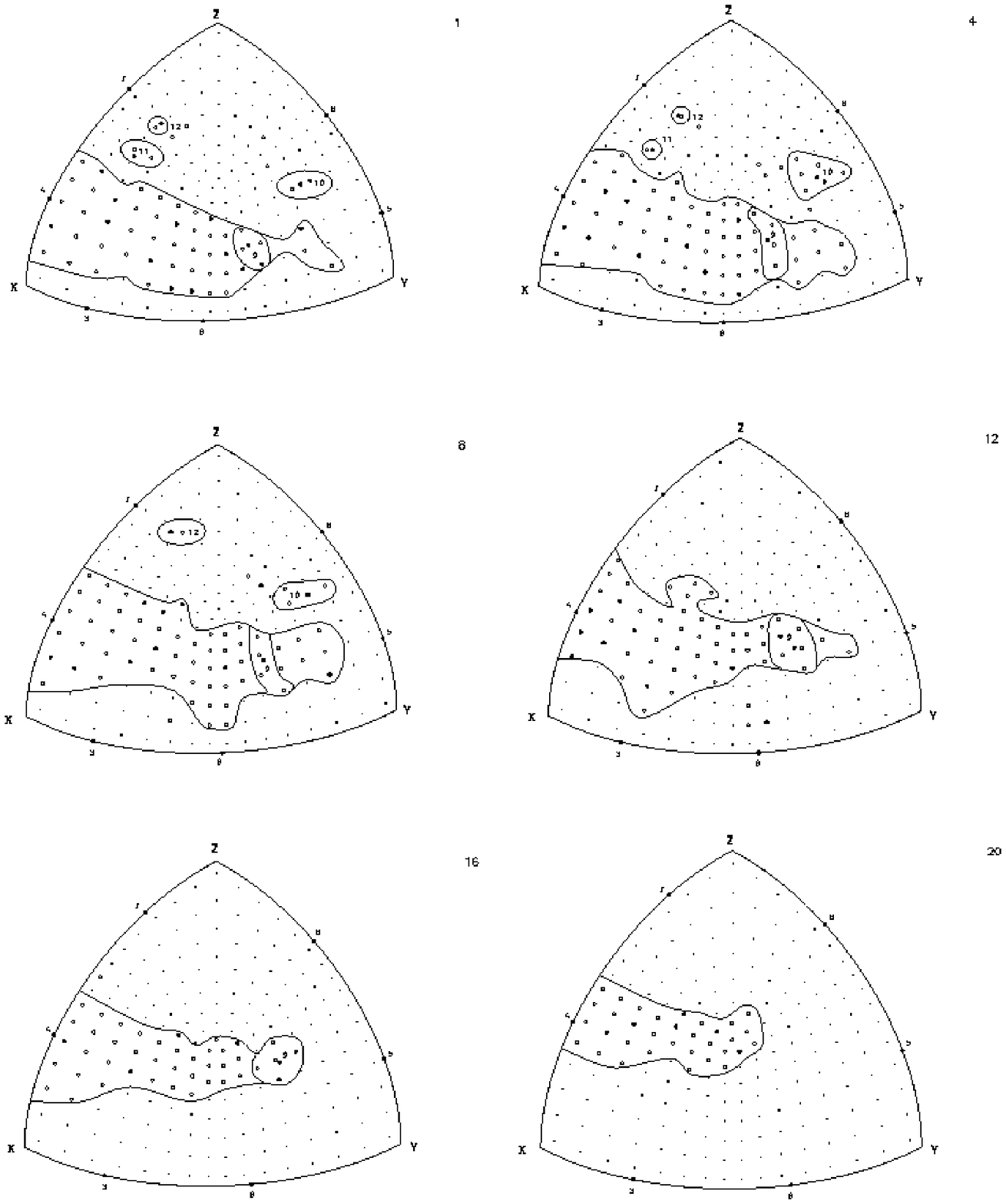}
{\normalbaselineskip=8pt\normalbaselines\noindent
Figure 6. Like Figure 4, for the strong-cusp model.
Resonant orbits are listed in Table 5.
\par}

\vfill\eject

\noindent
As we will see, the greater variety of regular orbit types in the
weak-cusp model makes it easier to find a self-consistent
solution that does not strongly weight the stochastic orbits.

Our classification scheme differs in one important respect from that
of Schwarzschild (1993), who assigned even some stochastic orbits to
families associated with stable periodic orbits (see his Figs. 4 and
5) - presumably on the
basis that many stochastic orbits behave like regular orbits over
short time intervals.
We chose instead to maintain a clear distinction between regular and
stochastic orbits, for the reasons outlined above.
In practice, of course, some misclassifications will always occur.

\subsect {Stochastic Orbits}

A large fraction of the orbits at all energies from the
stationary start space, and a
significant fraction of the orbits from the $X-Z$ start space, were
found to be stochastic in both model potentials.
Figure 7 shows the fraction of stochastic orbits as a function of
shell number for both models.
We emphasize that these fractions can not be simply related to
phase-space volumes; nevertheless they give a crude picture of the
relative importance of stochasticity at different energies.
In the weak-cusp model, the stochastic fraction is $\sim 0.6$ at low
energies in the stationary start space and increases to $\sim 0.8$
at high energies.
In the strong-cusp model, this fraction is relatively
constant with energy at $\sim 0.8$.
The fraction of stochastic orbits in the $X-Z$ start space is
lower in both models: roughly $0.1$ - $0.2$ at low energies,
falling to zero at high energies.
We interpret the much lower fraction of stochastic orbits from the $X-Z$
start space in terms of the smaller perturbing effect of the central cusp:
orbits begun
from the $X-Z$ plane are mostly tubes, which avoid the center.

Remarkably, the overall fraction of stochastic orbits is approximately
the same in the two models, even though the strength of the cusp -
which is presumably responsible for generating much of the
stochasticity - is very different in the two potentials.
However the numerically-computed Liapunov exponents are smaller by a
factor of a few (after scaling by the dynamical time)
in the weak-cusp model than in the strong-cusp
model.
Thus the stochastic orbits in the weak-cusp model should take longer to
diffuse through phase space than those in the strong-cusp model.
We now discuss the consequences of these different instability rates
for the behavior of stochastic orbits over longer time scales.

{\epsfxsize 5.5in \epsfbox{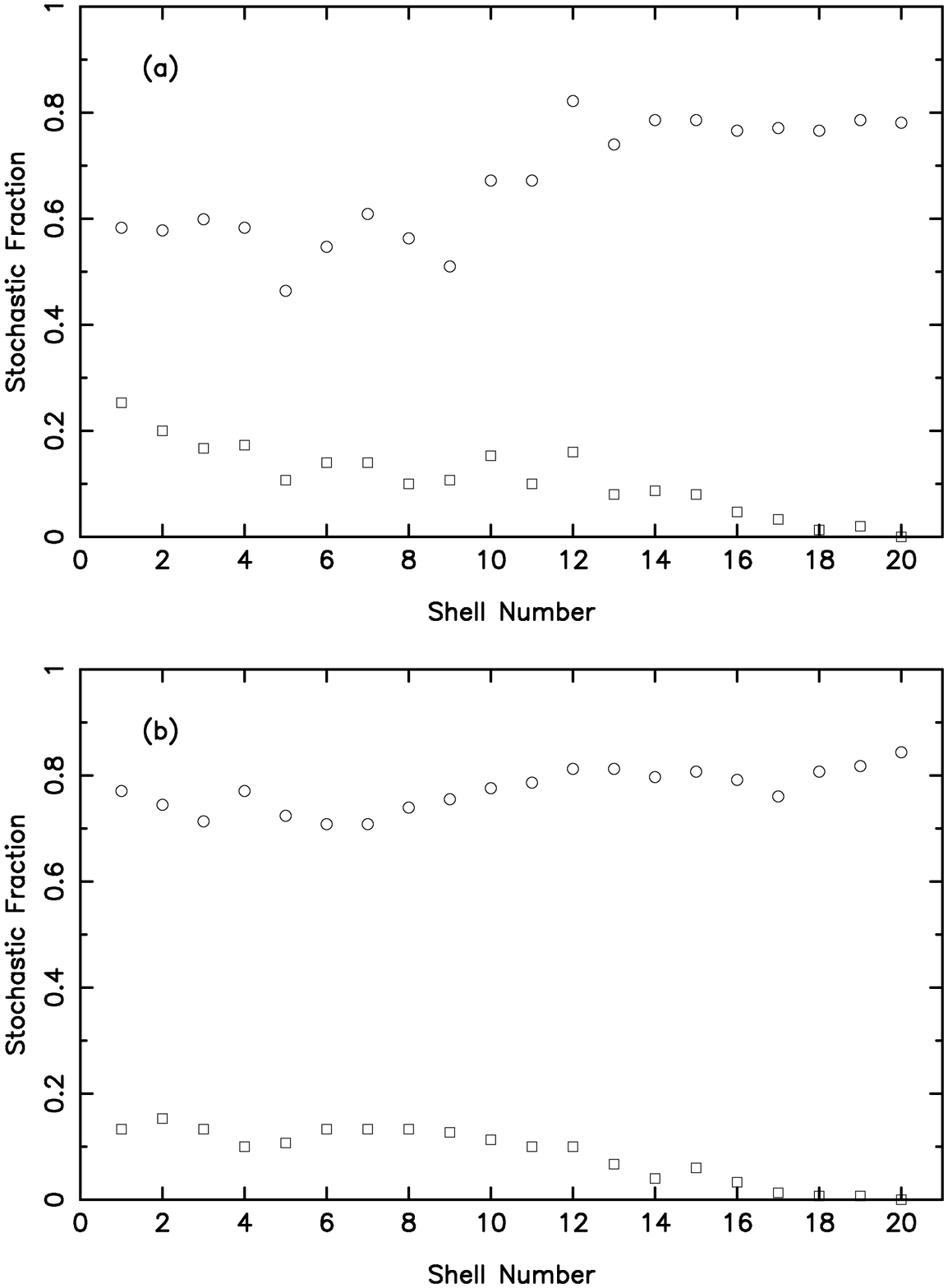}
{\normalbaselineskip=8pt\normalbaselines\noindent
Figure 7. Fraction of stochastic orbits per shell in the orbit
libraries, in the weak-cusp (a) and strong-cusp (b) models.
Circles: stationary start space; squares: $X-Z$ start space.
\par}

\bigskip

It is generally believed that the stochastic trajectories of a given
energy in a system
with three degrees of freedom are connected into a single complex
network, the so-called ``Arnold web'' (e.g. Lichtenberg \& Liebermann
1992, p. 380 ff).
For any initial condition within the stochastic region, a trajectory
will eventually come infinitesimally close to every point in phase space
consistent with energy conservation.
This diffusion is sometimes broken into two parts: a relatively fast
diffusion, resulting from the essentially random motion of the
phase point throughout a stochastic layer
surrounding an unstable periodic orbit; and
a slow diffusion, the ``Arnold diffusion,'' that links together parts
of stochastic phase space that are nearly isolated by the presence of
invariant tori associated with regular orbits (Arnold 1964;
Lichentenberg \& Lieberman 1992, p. 386).
The time scale for the fast diffusion can be as short as a dynamical
time, particularly if the phase space is globally stochastic so
that the trajectory wanders essentially randomly over the entire
energy surface.
An example is the H\'enon-Heiles (1964) potential at high energies.
The slower, Arnold diffusion is usually associated only
with systems of three or more spatial dimensions, where the topology
of phase space implies that the stochastic trajectories are fully
interconnected.
Arnold diffusion takes place on a
time scale that can be arbitrarily long if the phase space is
nearly integrable (Chirikov 1979), and it is usually assumed to be of
negligible importance in stellar systems which have lifetimes of only
$10^2-10^4$ dynamical times.

The qualitative nature of the diffusion seen in the present models
has been discussed by a number of authors,
especially Gerhard and collaborators (Gerhard \& Binney 1985;
Gerhard 1987), who showed that box orbits are destabilized by close
passage to a central mass concentration.
While the angle of deflection produced by even a $1/r^2$ density cusp
is generally small (Miralda-Escud\'e \& Schwarzschild 1989),
the dependence of this angle on
pericenter distance is large, leading to a sensitive dependence of the
orbital trajectory on initial conditions.
A stochastic orbit in such a potential may look similar to a
regular, box orbit for a
limited time, but will eventually undergo a scattering event that moves the
phase point to a different box orbit, etc.
After a sufficiently long time, any such trajectory would be expected
to uniformly fill the phase space region defined by all orbits that
pass near the center at that energy (Gerhard \& Binney 1985).
The result is a time-averaged configuration-space density that
is similar to, though rounder than, that of a typical box
orbit.

Gerhard \etal's numerical experiments in two spatial dimensions
(see also Binney 1982a) show that this diffusion can be relatively fast,
producing significant changes in the appearance of an orbit over just
tens or hundreds of oscillations if the central mass concentration is
sufficiently strong.
The diffusion described by Gerhard and Binney can not
be identified with Arnold diffusion, since Arnold diffusion
does not exist in systems with only two degrees of freedom.
Instead we associate it with the faster diffusion that occurs in a
strongly chaotic part of phase space.
We likewise assume that the orbital evolution seen in our
three-dimensional models is best associated with this faster
diffusion.
Over much longer time scales, Arnold diffusion might lead to
qualitatively different behavior for some orbits --- we can say
nothing useful about this based on our short integrations.

In our weak-cusp potential, inspection of the orbit plots reveals
that most stochastic trajectories do in
fact mimic regular box orbits over 100 dynamical times:
stochastic orbits of a given energy, but started from different
initial points, retain
distinctly different shapes over this limited period of time.
In the strong-cusp potential, on the other hand,
stochastic orbits quickly sample most of the
configuration-space volume contained within the equipotential surface and
attain a characteristic,
roughly spherical shape that is distinctly different from that of
a thin box orbit.
Thus the stochastic orbits in the strong-cusp potential appear to diffuse
over their allowed phase space volumes much more quickly than in the weak-cusp
model -- a result that is consistent with (though not strictly
implied by) the generally larger
Liapunov exponents of orbits in the strong-cusp model.

The central parts of many early-type galaxies are much older than 100
dynamical times.
Using the parameters given in Table 4 of Lauer \etal~ (1995), we
can compute approximate dynamical ages for ``power-law''
galaxies observed with HST.
(We assume spherical symmetry, circular orbits and M/L = 10 in
solar units.)
Defining a galactic lifetime as $5\times 10^9$ years, we find that
stars in NGC 3115 would have completed $10^3$ orbits at a radius
of 540 pc (=0.20$R_e$); $10^4$ orbits at 43 pc (0.016$R_e$); and
$10^5$ orbits at 3.3 pc ($1.3\times 10^{-3} R_e$).
In NGC 7332, we find dynamical ages of $10^3$ at 450 pc
(0.13$R_e$); $10^4$ at 41 pc (0.012$R_e$): and $10^5$ at 3.6 pc
($1.1\times 10^{-3}R_e$).
Both of these galaxies are luminous ($M_B\approx -19.5$);
lower-luminosity galaxies like M32 are typically denser and a
larger fraction of the mass of such a galaxy would have undergone
$10^3$ or more orbits by now.
Thus the behavior of a stochastic orbit
during $10^2$ dynamical times may not be a sufficient basis on which
to build a model that will maintain its shape
over the age of the universe.

We therefore investigated how the diffusion of the stochastic orbits
affects their time-averaged density distributions over time scales longer
than 100 periods.
We integrated particular stochastic orbits for up
to $10^5 T_D$ in our two model potentials.
To save computer time, the Liapunov exponents were not computed.
Instead, the time-averaged occupation numbers in a grid of cells
were recorded at
30, 100, 300, 1000, 3000, $10^4$, $3\times 10^4$ and $10^5$ orbital times.
(The grid of cells was constructed as described in \S 5, except that the
radial grid boundaries were equipotential surfaces, not
equidensity surfaces.)
In addition, we computed as a function of time a heuristic measure of the
departure of the orbit from ``ergodicity on the energy surface,''
defined as follows.
\footnote{$^1$}
{We follow the standard usage here, i.e. a ``stochastic'' or
``chaotic''
trajectory is one with non-zero Liapunov exponents, and an ``ergodic''
trajectory is one that uniformly fills some phase space volume --
either the entire energy surface, or some allowed part of it -- after
infinite time.
Ergodicity does not imply stochasticity; for instance, the motion of a
regular orbit is ergodic over its invariant torus.
Nor does stochasticity necessarily imply ergodicity, although
motion in a fully stochastic phase space can sometimes be
shown to be ergodic (Sinai 1976).
Our usage differs from that of some authors, e.g.
Goodman \& Schwarzschild (1981) use the term ``stochastic'' to
mean ``filling the energy hypersurface.''}
In an imaginary potential characterized by no integrals of motion aside
from the energy, every trajectory would be free to wander over every part
of the energy surface $E=E_0$.
Although the detailed trajectory of such an orbit is not
possible to calculate, its asymptotic, time-averaged configuration space
density would be
$$\eqalignno{
\rho_{erg}({\bf x})&\propto \int\delta(E-E_0)\ d^3 v &(\new a) \cr
&\propto\int\delta(E-E_0)\sqrt{E-\Phi({\bf x})}\ dE &(\last b) \cr
&=C\sqrt{E_0-\Phi({\bf x})}, \ \ \ \ \ \Phi({\bf x})\le E_0;&(\last c)\cr
&=0,\phantom{\sqrt{E_0-\Phi({\bf x})}}\ \ \ \ \ \ \Phi({\bf x})\ge E_0.\cr}
$$
Clearly we cannot expect any stochastic orbit in our model potentials
to attain this density distribution, even after an infinite time, since no
stochastic orbit can sample the regular parts of phase space.
Nevertheless we might expect the time-averaged density of
a stochastic orbit to approach $\rho_{erg}$ fairly closely,
particularly in models -- like ours -- where the fraction of
phase space associated with stochastic orbits is large.

As a measure of the deviation of the time-averaged density of a
stochastic orbit from this ``fully ergodic'' density, we defined
$$\eqalignno{
\Delta(t) &= \int\left[{\rho_{erg}({\bf x}) -
\overline{\rho}({\bf x};t)\over\rho_{erg}({\bf x})}\right]^2\rho_{erg}({\bf
x}) \ d^3x, &(\new a) \cr
&= \int\left[{\overline{\rho}({\bf x};t)\over
\rho_{erg}({\bf x})}\right] \overline{\rho}({\bf x};t) d^3x - 1 &(\last b)\cr}
$$
with $\overline{\rho}({\bf x},t)$ the time-averaged density of the
stochastic orbit at time $t$; the second expression follows after
assuming that $\rho_{erg}$ and $\overline{\rho}$ are normalized
to unit total mass.
Although $\overline{\rho}({\bf x},t)$ is not a well-defined quantity
for a single trajectory, we can compute a coarse-grained approximation
by recording passages of the trajectory through a grid of cells; for
our purposes, a natural set of cells are the ones just described.
We then have:
$$
\Delta(t) \approx \sum_{cells}{\overline{m}^2(t)\over m_{erg}} - 1,
\eqno(\last c)
$$
with $\overline{m}$ the time-averaged mass of the orbit in one grid
cell, and $m_{erg}$ the mass of the ``fully ergodic'' orbit in that cell.
The parameter $\Delta(t)$ is zero if and only if the stochastic
orbit has
a time-averaged density that is everywhere identical to that of the
``fully ergodic'' orbit of the same energy.
A value of $\Delta=p^2$ indicates that the rms deviation of
the cell mass from that of the ``fully ergodic'' orbit is $p$.

{\epsfxsize 5.5in \epsfbox{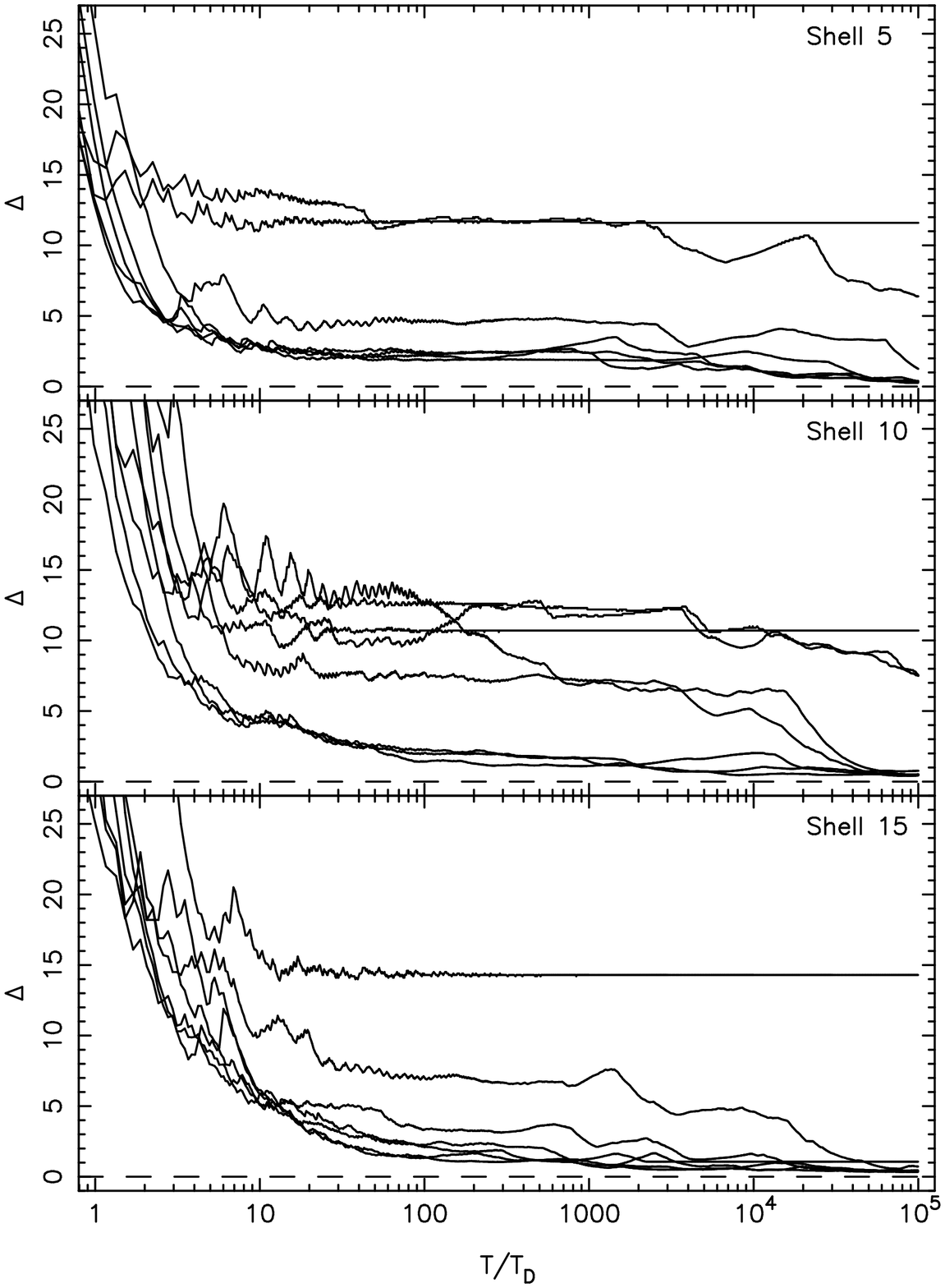}
{\normalbaselineskip=8pt\normalbaselines\noindent
Figure 8a. Approach to ergodicity for stochastic orbits in the
weak-cusp models.
\par}

\bigskip

{\epsfxsize 5.5in \epsfbox{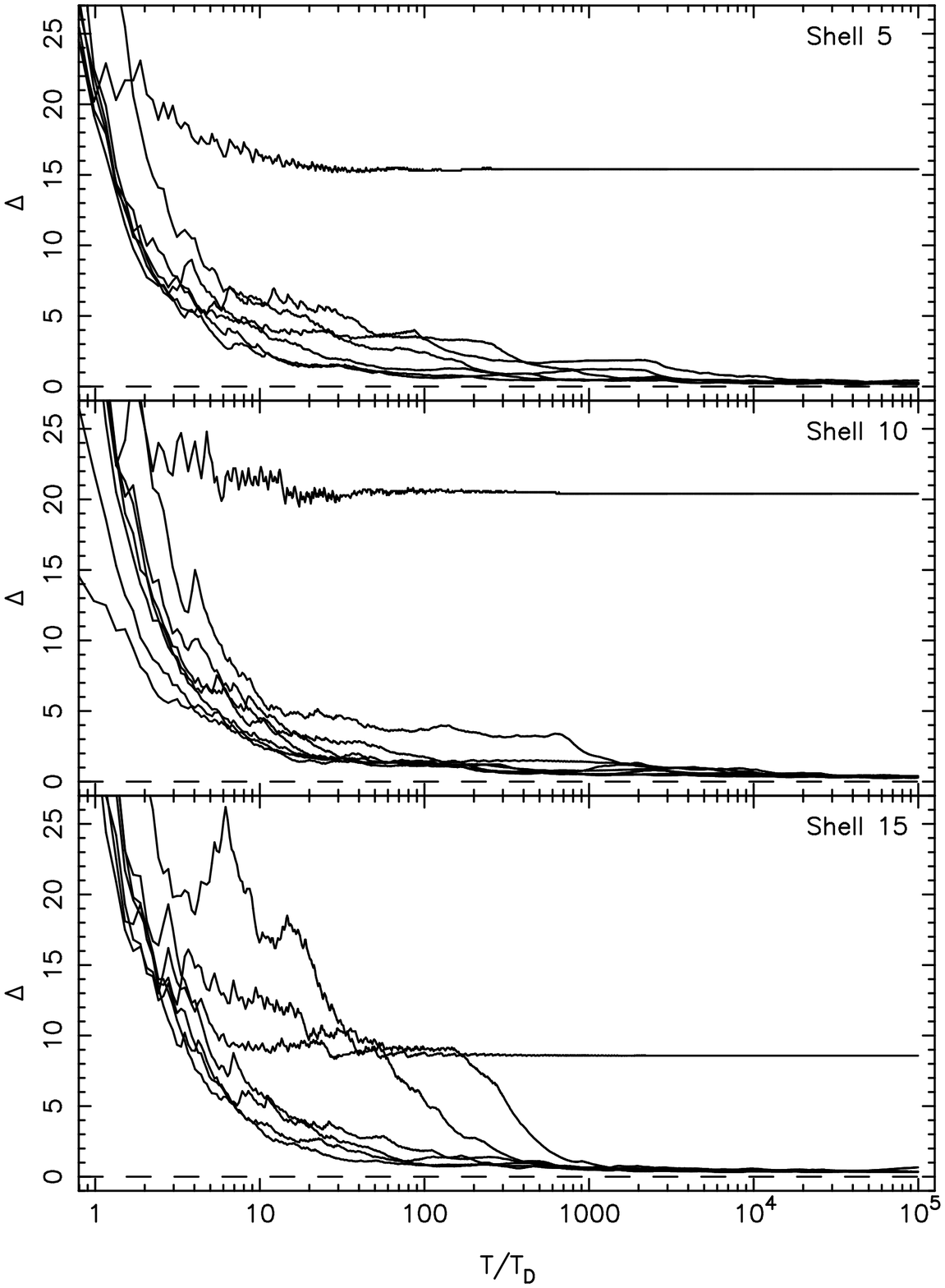}
{\normalbaselineskip=8pt\normalbaselines\noindent
Figure 8b. Approach to ergodicity for stochastic orbits in the
strong-cusp models.
\par}

Figure 8 shows the evolution of $\Delta(t)$ for a set of
stochastic orbits.
In each of the two model potentials, stochastic orbits from the
stationary start space were integrated at three
energies corresponding to shells 5, 10 and 15.
These orbits were chosen to have Liapunov exponents -- computed over 100 $T_D$
-- spanning the observed range for orbits at that energy.
The evolution of $\Delta(t)$ for one regular orbit at each energy is shown
for comparison.

Figure 8 illustrates the above-mentioned difference between
the stochastic orbits in the two potentials.
In the strong-cusp potential, all of the stochastic orbits evolve
rapidly toward a well-defined, time-averaged state that is
similar to that of the fully ergodic orbit of the same energy, and very
different from that of a typical regular orbit.
Even after $\sim 10^2$ oscillations, most stochastic orbits have
spatial distributions that are closer to each other than
they are to a typical regular orbit -- consistent with the
conclusions reached on the basis of the orbital plots.
By $10^3$ or $10^4$ dynamical times, the time-averaged densities
of the stochastic orbits are almost
indistinguishable from one another, though they remain distinctly
different from that of the fully ergodic orbit,
with $\Delta\approx 0.3\pm 0.05$.
Thereafter, the time-averaged densities remain nearly constant.

By contrast, the stochastic orbits in the weak-cusp potential are
generally
slower to fill their allowed phase-space volumes, and some
apparently never do.
After $10^2$ orbital times, there is as much spread in the $\Delta$
values among the stochastic orbits as between stochastic and regular
orbits; and even after $10^4$ or $10^5$ dynamical times the
time-averaged densities
of some of the stochastic orbits are showing little
tendency to evolve toward a steady state (although a few appear
to be nearly as ``ergodic'' as the orbits in the strong-cusp
potential).
A typical stochastic orbit in this potential appears to remain confined
to a restricted region of phase space over hundreds or thousands
of oscillations at least; thus many of the stochastic orbits
in this model can mimic regular orbits for
periods of time that are comparable to the age of the universe.
This result is consistent with that of other workers who
found that the stochastic orbits in nearly-integrable potentials
are far from ergodic over the energy surface on short time scales
(e.g. Goodman \& Schwarzschild 1981).

Figure 8 suggests that -- in at least the strong-cusp model --
we can define a single time-averaged density
distribution that represents every stochastic orbit at a given
energy.
This time-averaged density is illustrated in Figure 9 for shell 10;
the cell occupation numbers are averages from six
stochastic orbits, each integrated for $10^5$ dynamical times.
(Each individual stochastic orbit produced a time-averaged distribution that
differed only slightly from this average.)
Also shown for comparison is the density distribution of the ``fully
ergodic'' orbit, equation (17), of the same energy, as well as a
``boxlet'' from the 4:5:7 resonant family.
The stochastic and fully ergodic density distributions are
similar in the sense that all cells lying within the
equipotential surface are occupied.
However the stochastic orbit spends proportionately more time near
that surface.
In addition, the stochastic orbit is more highly flattened:
the occupation numbers of cells near the $x$-axis
are relatively higher than those of the fully ergodic orbit
when compared to cells near the $y$- and $z$-axes.
This flattening of the stochastic orbit can be understood
in one of two ways.
On the one hand, the stochastic trajectory may be seen as a
superposition of regular box orbits, as described by Gerhard \& Binney
(1985).
Since box orbits are

{\epsfxsize 5.5in \epsfbox{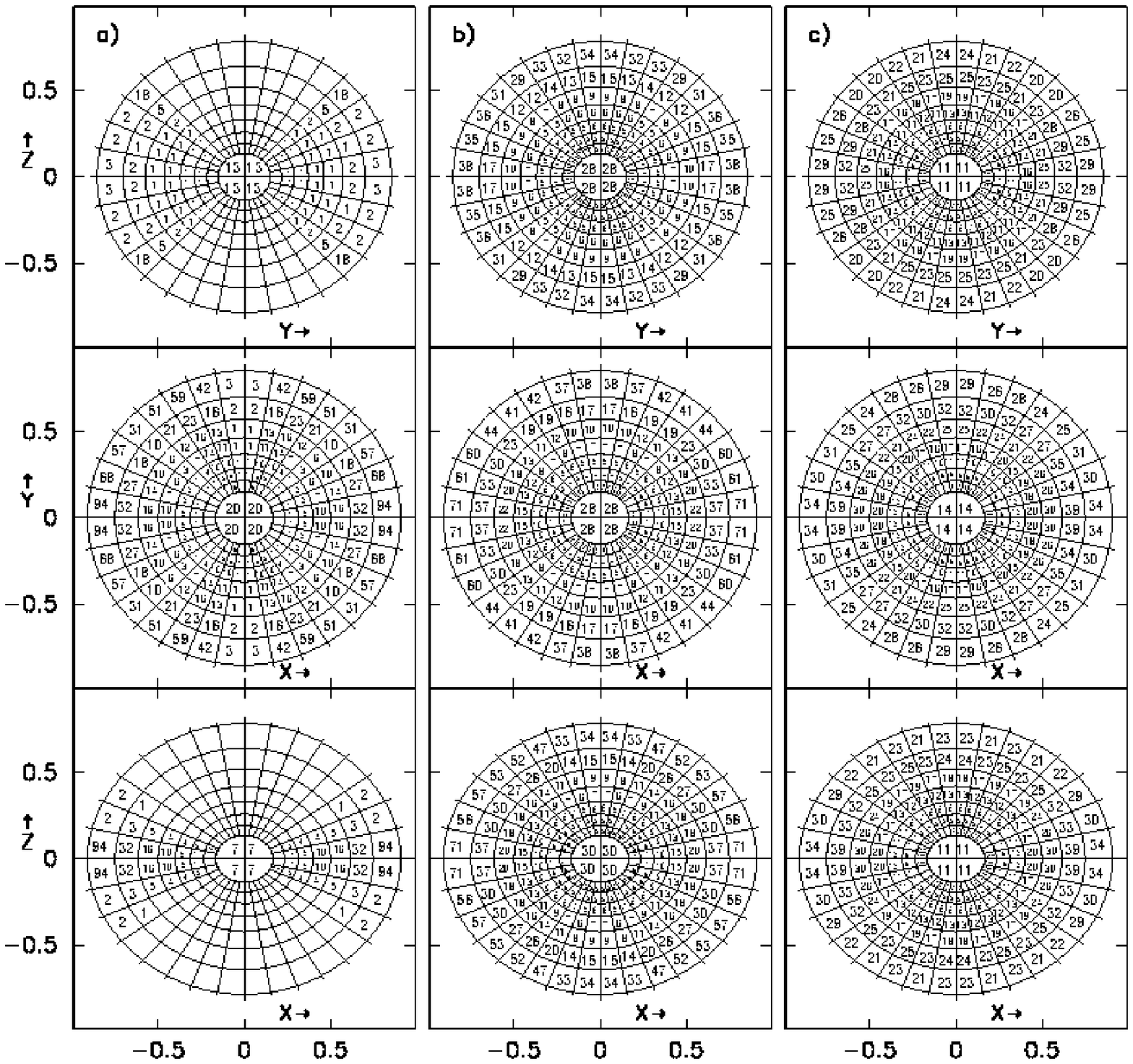}
{\normalbaselineskip=8pt\normalbaselines\noindent
Figure 9. Three cross sections through a regular orbit (a), a
time-averaged stochastic orbit (b) and a ``fully ergodic orbit''
(c) from shell 10 in the strong-cusp model.
The number in each cell represents the fraction of time (scale
arbitrary) which the orbit spends in the cell (numbers were
replaced by dots in some of the inner shells).
Cells without numbers are not entered by the orbit.
\par}

\vfill\eject

\noindent
primarily aligned along the $x$-axis, it is not
surprising that the time-averaged trajectory of a stochastic
orbit shows a similar elongation.
On the other hand, we can think of the stochastic trajectory as
similar to the fully ergodic trajectory minus those parts of phase
space that are regular.
The regular parts of phase space consist mostly of tube orbits, and
orbits from the tube families tend to enhance the density more
parallel to one of the shorter two axes than parallel to the
longer two axes.
Excluding these parts of phase space from the regions visited by the
stochastic trajectory would again imply a time-averaged density that
is elongated along the long axis.

These results suggest that the stochastic orbits in models with
strong cusps can explore their full phase space volumes over time scales of
order $10^2 -- 10^3$ dynamical times.
Thus -- at least in the inner regions of a model where the
orbital times, scaled to a real galaxy, would fall below
$\sim10^{-3}$ of the age of the universe -- we must replace all the
stochastic orbits of a given energy by a single, time-averaged orbit
that represents a uniform density of stars within the stochastic
part of phase space.

In models with a weak cusp, where diffusion time scales are
longer, one might be justified in assigning arbitrary occupation
numbers to ``different'' stochastic orbits of the same energy,
each integrated for a finite time.
However Figure 8 shows that many stochastic orbits even in the
weak-cusp potential behave nearly ergodically over modest time
scales, and a self-consistent solution that includes
distinguishable stochastic orbits would therefore not represent a
true equilibrium.
Furthermore one can imagine physical processes that might enhance
the diffusion rate above what is seen in Figure 8.
We return to these questions in \S 5.

We note that the estimation of phase-space filling rates via time-averaging
is not strictly justified, since nature is more
concerned with the evolution of ensembles of orbits than with time-averages
of individual orbits.
For instance, a regular orbit fills its torus in a time-averaged sense
within a few tens of crossing times, as shown here, but an ensemble of
points on a given torus {\it never} reaches an equilibrium distribution --
the ensemble simply translates, unchanged, around the torus.
The physically more interesting timescale is the ``mixing'' time,
i.e. the time required for an initially non-random distribution of
stars in phase space to mix into a time-invariant state.
Regular orbits, unlike chaotic orbits, do not mix, at least in a fixed
potential, and it is unclear how to define mixing rates for
these orbits in a manner that allows a meaningful comparison with
chaotic orbits.
A fuller discussion of these issues may be found in Merritt (1996).

\subsect{Comparison with Earlier Work}

The results described in this section are surprising in at least
one respect.
Earlier studies of the dependence of stochasticity on cusp strength
(e.g. Gerhard \& Binney 1985; Gerhard 1987) suggested that while
stochasticity would be
important in a triaxial model with a strong cusp,
$\rho\propto 1/r^2$ or steeper, the orbital
population of a model with a cusp as weak as $\rho\propto
r^{-1}$ would be essentially similar to that of a fully regular model.
These studies were mostly based on surfaces of section describing
the two-degrees-of-freedom motion in one of the principal planes.
In the present study (based admittedly on a single choice for the axis
ratios of the figure) we reach a very different conclusion: the fraction of
stochastic orbits is quite large even for weak cusps.
How can this discrepancy be understood?

Discussions with O. Gerhard and D. Pfenniger led to the
following insights.
Surfaces of section based on motion in the principal planes can be
misleading about the full, three-dimensional motion.
This is because, first, periodic orbits which generate families of
regular orbits in the principal planes are often unstable to
perturbations out of the plane;
second, there are
families of orbits that exist entirely outside of the
principal planes and so make no appearance on the surfaces of section;
and third, the time scale required for an orbit to manifest its
stochasticity may be rather longer than the typical integration time
used in constructing a surface of section.

We constructed surfaces of section in the three principal planes of
our models and found that the vertical instability of the planar
periodic orbits is the most important of these three factors.
Only the $x-z$ bananas, among all the low-order planar periodic
orbits, is stable at most energies to vertical perturbations in our models
(Tables 3 and 5).
Thus the surfaces of section in the $x-y$ and $y-z$ planes
show large regions of regular motion
associated with the banana, fish etc. orbits, while in fact this regularity
is destroyed once the motion is perturbed out of the plane.
Indeed, based on the surfaces of section, it is the $x-z$ plane
that
contains the {\it greatest} amount of stochastic motion (generated by the
unstable $x$ and $z$-axial orbits), while in fact the 3-D motion in
the vicinity of this plane is much {\it more} regular than around the
other two due to the vertical stability of the $x-z$ bananas.

\sect{Construction of Self-Consistent Models}

Self-consistent solutions were constructed in the usual manner
(Schwarzschild 1979).
The fraction of time spent by each orbit in a grid of cells was
recorded, then a linear superposition of orbits was
sought, with non-negative occupation numbers, that reproduced the
known mass of the model in each cell.

The grid of cells was defined as follows.
As described above, the models were divided by 20 ellipsoidal shells into 21
sections of equal mass.
Each octant of each section was then divided into 48 facets, in the
following manner.
The octants were first divided into three sectors by the planes
$z=cx/a$, $y=bx/z$ and $z=cy/b$.
Then each sector was divided into 16 facets by a set of six planes.
In the case of sector one, which contained the $x$-axis,
these planes were defined by $ay/bx=1/5,2/5,2/3$ and
$az/cx=1/5,2/5,2/3$.
The facets in the other two sectors were defined in a symmetric way.
This grid cell structure has the advantage that the mass which the
model places in each cell is roughly the same, and furthermore these masses
are precisely independent of the model axis ratios
(although they still depend on the density profile, as specified by $\gamma$).

The problem to be solved is a constrained optimization problem.
The quantity to be optimized (minimized) is the discrepancy
between the model cell
masses, and the cell masses generated by some linear combination of
the orbits; the constraints include, among other possible things,
the non-negativity of the orbit occupation numbers.

Various criteria have been adopted in the past for deciding when a
combination of orbits comes ``close enough'' to reproducing the model cell
masses.
Schwarzschild (1979) originally required his solutions to
reproduce each of the cell masses to within machine precision.
More recently (Schwarzschild 1993), he has required only that
the maximum error in one grid cell not exceed 1\%.

Clearly a requirement that the model densities be reproduced to within
machine precision is overly stringent given the other approximations that
are made in the construction of the model (finite integration times,
discrete grids, etc.)
Furthermore the problem that we are solving is ill-conditioned
(O'Sullivan 1986) in the sense that any attempt to
find a numerically exact solution of the discretized equations
will generate large-amplitude fluctuations in the weights from orbit to orbit.
Such an unsmooth solution is not necessarily a relevant model for a
real galaxy; a better goal would be to find a solution that is
smooth in phase space but which (as a consequence of the smoothness
constraint) might only reproduce the cell masses approximately.
A further argument against requiring exact self-consistency
is that failure to reproduce the cell masses exactly may only mean
that the orbit library is not sufficiently dense to cover the phase
space in a fully representative way.

We attempted to demonstrate self-consistency by giving larger and
larger subsets of our orbit libraries to the optimization routine, and
observing how the average error in the cell masses varied with orbit
number.
One might expect that -- if a self-consistent solution exists --
the error in the cell masses would decrease rapidly once the number of
orbits in the solution exceeded the number of grid cells, roughly
$10^3$.
On the other hand, a more gradual decrease of error with orbit number
would suggest the nonexistence of a self-consistent solution.
(Of course, our scheme does not rule out the possibility of reproducing
the cell masses exactly and in fact this sometimes occurred in
practice.)
We then attempted to construct smooth solutions, to verify that
the self-consistency was not an artifact of the discretization.

This scheme has the advantage of permitting a meaningful comparison between
solutions obtained using different subsets of orbits, e.g. all orbits,
regular orbits only, etc.
The disadvantage, of course, is that there may be cases where the error
falls neither so slowly nor so rapidly with number of orbits that
one can clearly decide whether a self-consistent solution exists.

We chose to minimize the mean square deviation in the cell
masses,

$$\eqnam{\chisq}
\chi^2 = {1\over N}\sum_{l=1}^N\left[D_l - \sum_{i=1}^M C_i
B_{il}\right]^2,
\eqno (\new)
$$
where $C_i$ is the number of stars on orbit $i$, $1\le i \le M$;
$B_{il}$ is the mass which the $i$th orbit places in the $l$th cell;
and $D_l$ is the mass which the model places in the $l$th cell.
Self-consistency was not sought in the outermost shell of grid cells,
since few orbits supplied densities in these cells.
Thus $N=912$ is the number of grid cells in the model excluding those in
the outer shell.

The basic set of constraints was
$$\eqnam{\constr}
C_i\ge 0, \eqno (\new)
$$
i.e. nonnegative orbit weights.
Other constraints are discussed below.

The constrained optimization problem represented by equations (\chisq)
and (\constr) is an example of quadratic programming.
We used the NAG routine E04NCF to carry out the optimization.
A solution using the full set of orbits required about six hours on a
DEC Alpha 3000/700 computer.

\subsect{Quasi-Equilibrium Solutions}

Not all solutions to the self-consistency equations presented above
represent fully stationary galaxy models.
The orbits in our libraries were integrated for only 100 dynamical
times, and the stochastic orbits do not
reach their invariant distributions in such a short time.
We can nevertheless attempt to construct ``quasi-equilibrium''
solutions
in which we treat the stochastic and regular orbits in the same way,
allowing each stochastic orbit to have an arbitrary
occupation number.
Such models would, if allowed to evolve, presumably
change shape near the center as the stochastic orbits
gradually filled phase space in a time-independent way.
Nevertheless they provide a useful starting point for considering
more fully stationary models.

Figure 10 shows the departure from self-consistency, as a function of
the number of orbits supplied to the optimization routine,
for the two mass models.
Orbits were chosen by selecting every 10th, every 5th, etc. orbit from
the complete libraries of 6840 orbits.
The departure from self-consistency was defined as
$$
\delta = {\sqrt{\chi^2}\over {\rm average\
mass/cell}},
$$
i.e. the percentage, rms deviation in the cell masses.
Figure 10 shows that both mass models admit
quasi-stationary solutions to the self-consistency equations.
The weak-cusp model yields $\delta=0$ when the number of orbits
exceeds $\sim 3000$, while the strong-cusp model shows a slight
departure from numerical self-consistency, $\delta\approx
2\times 10^{-4}$, even when the full set of orbits is used.
However Figure 10 strongly suggests that including a slightly
larger number of orbits in the strong-cusp solution would give
exact self-consistency for this mass model as well.

{\epsfxsize 5.in \epsfbox{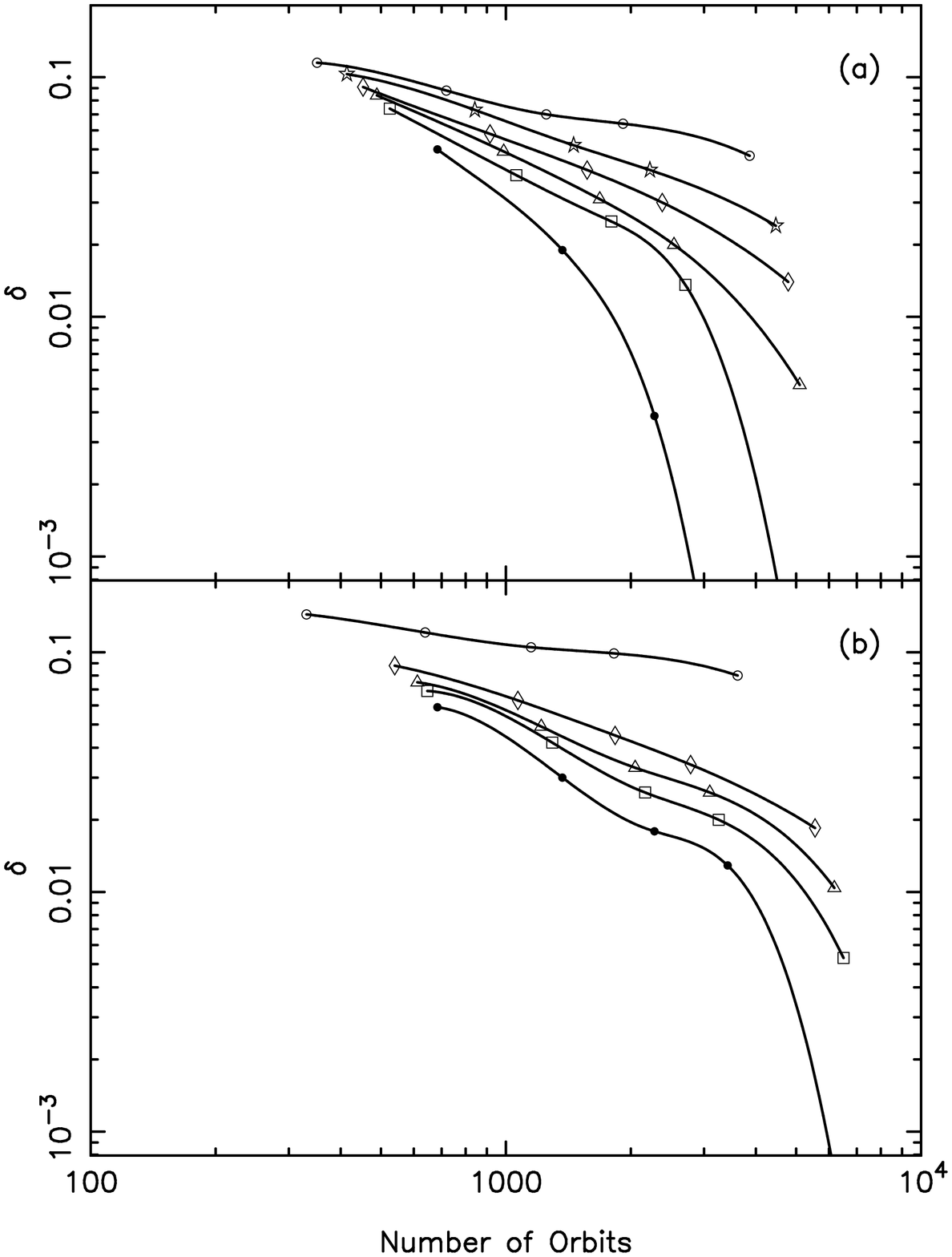}
{\normalbaselineskip=8pt\normalbaselines\noindent
Figure 10. Departure from self-consistency ($\delta$) as a
function of number of orbits supplied to the optimization
routine, for the weak-cusp (a) and strong-cusp (b) models.
$\circ$: regular orbits only; $\bullet$: all orbits.
The intermediate curves in (a) represent solutions in which the
stochastic orbits were included only in shells above 10
($\sqcap$), 12 ($\triangle$), 14 ($\diamond$), and 16 ($\star$).
In (b), the intermediate curves represent solutions in which the
stochastic orbits were included only in shells above 2
($\sqcap$), 4 ($\triangle$) and 8 ($\diamond$).
\par}

\bigskip

One (rather artifical) way to construct fully stationary solutions
is to supply only the regular orbits to the optimization routine.
Excluding all stochastic orbits leaves 3864 and
3616 regular orbits, respectively, in the weak- and strong-cusp
orbit libraries.
Figure 10 shows that the regular orbits alone can reproduce the
model cell densities with an accuracy of only $\delta\approx 5\%$
in the weak-cusp model and $\sim 8\%$ in the strong-cusp model;
furthermore the weak dependence of $\delta$ on orbit number does
not give confidence that a larger number of orbits would
greatly reduce the error.
We conclude that the regular orbits -- the tubes and boxlets -- do not
provide a wide enough variety of shapes to reproduce the triaxial
figure in either mass model.
However the regular orbits go farther toward reproducing the
model density in the weak-cusp case, presumably because of the
greater variety of boxlet families in this potential (Fig. 4).

Since the diffusion time scale for stochastic orbits is a strong
function of their energy, one might hope to construct a nearly
stationary model by simply excluding the stochastic orbits with
energies below some threshold.
The resulting self-consistent solution, if it exists, would depart
only negligibly from equilibrium at radii less than the typical radius
of the most bound stochastic orbits included.

Figure 10 demonstrates that such ``nearly equilibrium'' solutions
can in fact be found, at least for the weak-cusp model.
Removing all of the stochastic orbits inside
shell 10 --- roughly the inner half-mass radius --- does
not destroy the self-consistency of this solution.
Since the diffusion time scale for stochastic orbits outside of the
half-mass radius of a galaxy is likely to exceed $10^{10}$ years (\S 4),
we conclude that a galaxy constructed in this way would be
effectively stationary for a Hubble time.

In the case of the strong-cusp model, however, the deviation from
exact self-consis- tency increases rapidly as one excludes stochastic
orbits from larger and larger regions near the center.
Figure 10 suggests that only the stochastic orbits inside shell 1 or
2 can be excluded without destroying self-consistency.

Schwarzschild (1993) found that he could achieve nearly exact
self-consistency
in a scale-free, $\rho\propto r^{-2}$ triaxial model, with
$c/a=T=1/2$, using only the regular orbits.
Schwarzschild's model can be seen as an approximate representation
of the central parts of our strong-cusp model where the density
also varies as $\sim r^{-2}$.
Our new result is consistent with Schwarzschild's, but
suggests that self-consistency using just the regular
orbits becomes more difficult as one includes parts of the
model that are not scale-free.
One contributing factor is that the $x-z$ ``banana'' orbits --
which were found to contribute about 80\% of the mass not associated
with tube orbits in these fully-regular solutions -- become more
sharply bent at large energies, and thus less able to maintain the
strong flattening of these models at large radii.
Regular orbits alone might therefore have sufficed if our mass
model had been chosen to be more nearly round.
In addition, Figure 6 shows that -- while the fraction of
regular orbits is not a strong function of energy in the
strong-cusp model -- the regular orbits at high energies belong
primarily to just one family, the $x-z$ bananas.
At low energies, in the scale-free part of the potential,
a wider variety of regular orbit families are available for
building a model.

\subsect{Fully-Mixed Solutions}

The experiments just described demonstrate that one can construct
nearly-stationary triaxial models of galaxies with weak cusps
by including stochastic orbits only at high energies.
But nature would almost certainly not build galaxies
in this way, since it has no mechanism for
selectively placing stars on regular or stochastic orbits and would
therefore tend to populate these two parts of
phase space in rough proportion to their volumes.
Here we explore self-consistent solutions in which the stochastic
parts of phase space are strongly populated, but (unlike in the
solutions described above) in an approximately time-independent manner.

We begin by noting that Jeans's theorem can be generalized
to include the (generic) case of a galaxy in which not all
of the orbits are regular.
The first step is to divide  phase space into regular and stochastic
regions.
In the regular parts of phase space, orbits are characterized by two
isolating integrals in addition to the energy, and
the Jeans theorem in its usual form is valid: the steady-state
distribution function $f_R$ -- which must be single-valued and
constant along trajectories -- can depend on the phase space coordinates
only through the isolating integrals of motion, $f_R=f_R(E,I_2,I_3)$.
(Note that the integrals $I_2$ and $I_3$ are now local, not global,
invariants and so their definitions will change from
one regular region to another.)
In the stochastic parts of phase space, which are interconnected through
the Arnold web, strict constancy of $f$ along trajectories
requires that the
phase space density have a single value throughout the entire stochastic
region at every given energy; thus $f_S=f_S(E)$.
The penalty for violating this condition will only be severe if the
stochastic orbits undergo significant diffusion on time scales of
interest, however. \footnote{$^2$}{
Binney (1982b) expressed the view that Jeans's
theorem applies only to completely integrable systems.
His central points, with which we agree, are that ergodicity can only
be rigorously proved for regular orbits, and stochastic orbits
may take a very long time to uniformly fill their allowed
phase-space regions.
However Liouville's theorem guarantees that an {\it initially} uniform
distribution of points in any bounded phase-space region will remain
uniform forever,
and this is true whether the phase-space region is regular or stochastic;
the ergodicity of individual orbits is irrelevant.
}

A steady-state galaxy will therefore have a distribution function of
the form
$$\eqnam{\jeans}
f = \cases {f_R(E,I_2,I_3) &(regular phase space)\cr
f_S(E) &(stochastic phase space) \cr} \eqno (\new)
$$
where it is understood that the integrals
$I_2$ and $I_3$ are local, not global, invariants of the motion.
We will refer to a model that is described by equation (\jeans) as ``fully
mixed''.

It was shown above that the
diffusion time for stochastic orbits in the strong-cusp
potential was relatively short, of order $10^2 -- 10^3$ dynamical times.
This result suggests that fully-mixed models might be relevant to
at least the central parts of real galaxies with strong cusps.
But even when modelling the outer parts of a galaxy, or the
inner parts of a galaxy with a weak cusp,
where diffusion time scales are relatively long, it is still
interesting and appropriate to construct fully-mixed models, for the
following reasons.

1. If a fully-mixed solution
corresponding to a given mass model can not be
found, then any galaxy with the same mass distribution must be
slowly evolving due to the continued mixing of its stochastic orbits.
The non-existence of fully-mixed models thus has interesting
consequences for the dynamical states of real galaxies.

2. A number of physical processes can be imagined that would increase the
diffusion rate of stochastic orbits above what has been calculated
here; and there is no obvious mechanism that would decrease it.
For instance, many elliptical galaxies contain
deviations from ellipsoidal symmetry in the surface brightness
at the $\sim 1\%$ level, e.g. ``boxy'' or ``disky'' isophotes
(e.g. Nieto \& Bender 1989).
These distortions -- to the extent that they are reflected in the
gravitational potential -- might be expected to enhance stochasticity above
what is observed in perfectly ellipsoidal models like ours (Udry
\& Pfenniger 1988).
Star-star interactions in a dense nucleus would help to scatter
trajectories in phase space by providing random,
time-dependent force perturbations (e.g. Goodman \& Schwarzschild 1981).
Any additional central mass concentration, such as a massive
black hole or dense
stellar nucleus, would produce occasional, large-angle deflections in
orbits passing near the center (Gerhard \& Binney 1985).
Tidal forces from nearby galaxies might also provide a time-dependent
perturbing force for stars on high-energy orbits.
These extra sources of diffusion might be of rather different
importance in different galaxies, but each would act to enhance the
mixing of stochastic orbits above that observed in simpler models.

3. Rapid and chaotic changes in the gravitational potential during
the formation of a galaxy are thought to redistribute stars in
phase space on a time scale that is of order the
collapse time (Lynden-Bell 1967).
These fluctuations act equally on stars whether or not they lie
in regular parts of phase space (or will lie in such regions, after
the potential reaches a steady state); their net result is to
generate a nearly uniform population of stars on constant energy
surfaces.
We might therefore expect some galaxies to {\it begin} their lives in
highly mixed states.
The strength of this argument depends on the efficiency
of violent relaxation and it is likely to be only more or less correct
in any particular galaxy.
Nevertheless one could take the view that the ``most probable'' model
for a galaxy is one in which the stochastic parts of phase space
are uniformly populated.

\medskip

One way to construct numerical approximations to fully-mixed models would
be to carry out very extended integrations of orbits in the stochastic
parts of phase space at every energy and record their time-averaged
densities.
Such experiments were described in \S3.
If the time integrations were sufficiently long, then every such time-averaged
trajectory would approximate a uniform population of the
stochastic part of phase
space and could be treated like a regular orbit in the model
construction -- with the important difference that only
{\it one} such orbit would be available at each energy.

{\epsfxsize 5.in \epsfbox{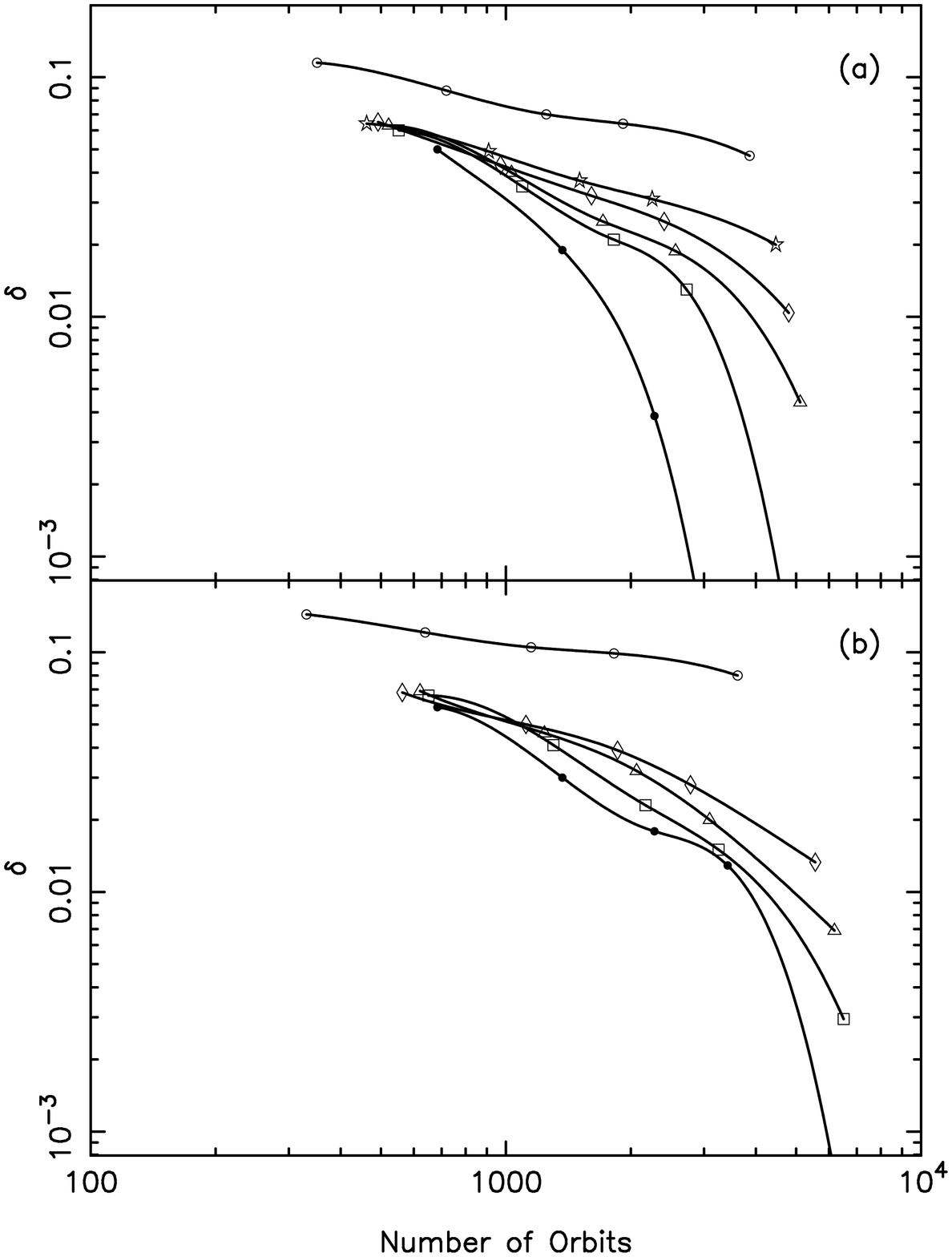}
{\normalbaselineskip=8pt\normalbaselines\noindent
Figure 11. Departure from self-consistency ($\delta$) as a
function of number of orbits supplied to the optimization
routine, for solutions in which the stochastic orbits
are ``fully-mixed'' below some energy.
(a) Weak cusp; (b) Strong cusp.
The intermediate curves in (a) represent solutions in which the
stochastic orbits were forced to be fully-mixed in shells below and
including 10 ($\sqcap$), 12 ($\triangle$), 14 ($\diamond$), and 16 ($\star$).
In (b), the intermediate curves represent solutions in which the
stochastic orbits were fully mixed in shells below and including 2
($\sqcap$), 4 ($\triangle$) and 8 ($\diamond$).
\par}

\bigskip

Following a suggestion of Ya. G. Sinai, we chose
instead to approximate these time-averaged densities by
simply arithmetically averaging the densities of the
$\sim 100$ stochastic orbits at
each energy in our orbit libraries, each already computed for the
standard 100 dynamical times.
In other words, we replaced a time average by an ensemble average --
a generally more efficient way to proceed, particularly in the case of
the weak-cusp model where the diffusion times are long.
We verified that these ensemble averages approximated closely the
direct time averages presented for a few stochastic orbits in \S 3.

One way to implement Prof. Sinai's suggestion is to simply include the
additional constraints
$$\eqnam{\newcon}
C_j=C_k \eqno (\new)
$$
into the quadratic programming routine, where $j$ and $k$ are understood
to refer to every pair of stochastic orbits of the same energy.
An equivalent, and computationally more efficient, approach is to
replace the stochastic orbits in the orbit
libraries by their arithmetic average at each energy before carrying out the
optimization.
The latter approach reduces the number of orbits that are supplied to
the quadratic programming routine and so increases its speed; we
followed it in the experiments described below.

The additional constraints represented by equation (\newcon) will make
it more difficult to find a self-consistent solution than if each
stochastic orbit were allowed to have its own weight, as in the
quasi-equilibrium solutions described above.
On the other hand, the time-averaged stochastic orbits (Fig.
9) are
elongated in approximately the same sense as the model figure, and
including them in the orbit library gives the optimization routine
slightly more flexibility -- in the form of just a single extra
``orbit'' at each energy -- than it would have if only the
regular orbits were included.

As Figure 11 shows, this slight additional freedom allows the quadratic
programming routine to reduce the residuals somewhat compared to
the solutions which exclude the stochastic orbits completely below
some energy.
For instance, in the strong-cusp model, $\delta$ drops by a
factor of $\sim 2$ when the fully-mixed stochastic orbits are
added to the regular orbits in the innermost shells.
Less improvement is seen in the weak-cusp solutions --- here, the
regular orbits provide a larger fraction of the mass near the
center.

Although the fully-mixed stochastic orbits are not a great asset
from the point of view of constructing a self-consistent model, are they
a serious liability?
In other words, could a galaxy place a large number of stars on such
orbits without destroying the self-consistency?
To investigate this question, we attempted solutions in which
the weights assigned to the time-averaged stochastic orbits
were increased as far as possible without violating
self-consistency.
Figure 12 shows one example for each mass model.
The weak-cusp solution shown there, which is completely self consistent
($\delta=0$), was constructed by maximizing the
numbers of stars on fully-mixed stochastic orbits from shells
1 through 6; stochastic orbits with larger energies were allowed
to have arbitrary occupation numbers.
The fully-mixed stochastic orbits carry approximately one-third of
the mass near the model center, while at larger energies,
stochastic orbits -- no longer fully mixed -- dominate.
The ``boxlets'' are important
only at low energies, where they likewise contribute about one-third
of the mass; at high energies their contribution is negligible.
Overall, stochastic orbits contribute 45\%, and
boxlets only 11\%, of the total mass in this solution.

The strong-cusp solution in Figure 12 contains fully-mixed stochastic
orbits at only the two lowest energies; we find $\delta =
3.3\times 10^{-3}$, nearly self-consistent.
Once again the stochastic orbits are dominant at almost every
energy, contributing 60\% of the total mass; boxlets carry only
4\%.

We conclude that a substantial population of stars distributed
uniformly throughout
stochastic phase space at low energies is consistent with
self-consistency in our weak-cusp model.
In the strong-cusp model, only the cusp itself can be so
populated without violating self-consistency.

\subsect{Smooth Solutions}

One disadvantage of an orbit-based approach to model building is that
the solutions are extremely unsmooth.
One source of this lack of smoothness is the discrete way
in which phase space is sampled.
But even more important is the inherent ill-conditioning of the
self-consistency problem.
A single orbit, which represents a delta-function in integral space,
covers a finite region in configuration space.
Deriving the integral-space density from the configuration space
density is therefore in the nature of a deconvolution problem, and
deconvolution has the property of amplifying errors or
incompleteness in the ``data'' (in our case, the masses which the
models place in the cells).
Even a highly noisy set of orbital weights can generate a smooth
configuration space density, and there are many more noisy
solutions than smooth ones.
This effect actually becomes {\it worse} as the number of orbits is increased,
since a fine grid is better able than a coarse grid to represent high-frequency
fluctuations (Phillips 1962).
In the absence of the non-negativity constraints, we would therefore
expect our solutions to exhibit wildly variable occupation numbers
from orbit to orbit, with many of the occupation numbers negative.
Even enforcing positivity should only reduce, not eliminate, these
fluctuations by ``clipping'' the negative occupation numbers at zero.
Indeed, many or most of the orbital weights in the solutions described above
were found to be zero, even when the number of orbits supplied to the
optimization routine was smaller than the number of grid cells in
which the mass was specified.
\footnote{$^3$}{Schwarzschild (1993) notes that his Richardson-Lucy
algorithm often required a
very large number of iterations, $\sim 4\times 10^4$, before reaching
approximate self-consistency.
This is a possible indication that his solutions were extremely
unsmooth, since Lucy's method always iterates toward an increasingly noisy
solution.
It may also imply that self-consistency was inconsistent
with smoothness for his models, since a smooth solution would
presumably have been reached more quickly if it had existed.}

Since our mass model is completely smooth, we might reasonably require
the phase-space density of our self-consistent solutions to be smooth
as well.
An unsmooth solution could be acceptable, but only if it can be
shown to have average properties that are close to those of a smooth solution.
If, however, imposing smoothness on a numerical solution causes it to
depart strongly from self-consistency, one would conclude that
no solution continuous in the phase-space density exists, and
that the apparent self-consistency is a numerical artifact
associated with the discretization.
It is primarily to address this concern that we now consider a
modified algorithm that generates smooth solutions.

{\epsfxsize 4.75in \epsfbox{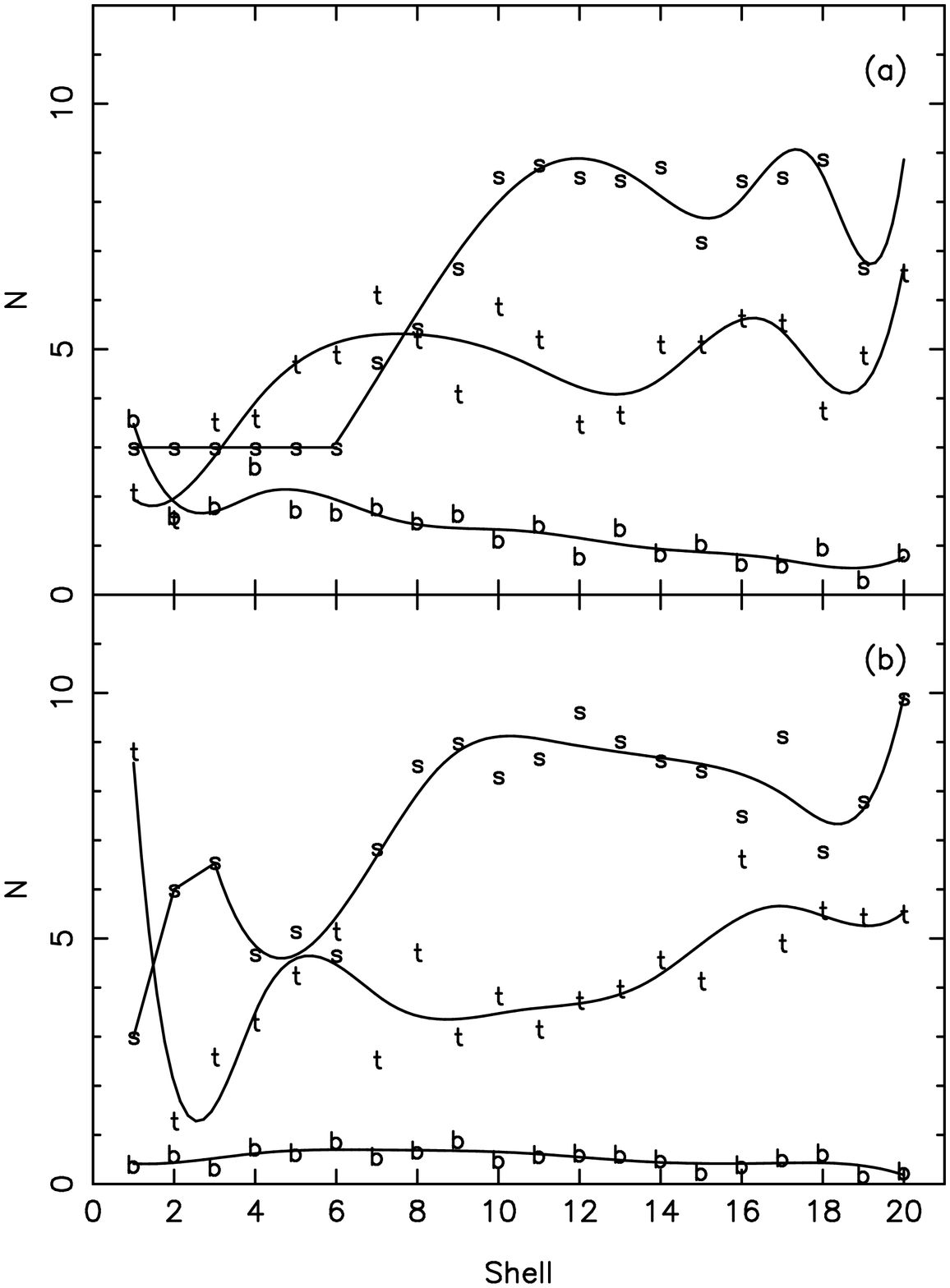}
{\normalbaselineskip=8pt\normalbaselines\noindent
Figure 12. Orbital content as a function of shell number for a weak-cusp
(a) and strong-cusp (b) solution.
In (a), the solution is fully mixed in shells below and including
6; in (b), the solution is fully mixed in shells 1 and 2.
The symbols $s$, $t$ and $b$ denote stochastic, tube and boxlet
orbits respectively.
The vertical axis is the number of stars from the indicated shell
on orbits of the three types (arbitrary normalization).
\par}

\bigskip

A proper prescription for smoothness in our case is that the density be
continuous in action space.
This is a difficult prescription to apply, since our phase space is
very complex and we have not computed the
actions associated with even the regular orbits.
Instead we will require simply that the orbital weights vary more-or-less
smoothly over our initial condition grids: that is, that
$$\eqnam{\con2}
C_j\approx C_k, \eqno (\new)
$$
where $j$ and $k$ represent two orbits with adjacent initial conditions.
Following Phillips (1962) and Tikhonov (1963),
the large number of constraints implied by
equation (\con2) can be imposed by adding a single term, a
``penalty function,''  to the optimization functional of equation
(\chisq).
The quantity to be optimized becomes
$$\eqnam{\opt2}
\chi'^2 = {1\over N}\sum_{l=1}^N\left[D_l - \sum_{i=1}^M C_i B_{il}\right]^2 +
\alpha \sum_{i=1}^M g(C_i)
\eqno (\new)
$$
subject to constraint (\constr) (nonnegative $C_i$'s).
We chose $g(C_i)$ to have the simple form
$$\eqnam{\entropy}
g(C_i) =  C_i^2.
\eqno(\new)
$$
This is ``zeroth-order regularization'' (e.g. Miller 1974), and has
the effect of filtering fluctuations on ``scales''
shorter than some maximum value determined by the smoothing parameter
$\alpha$.
(One might interpret (\entropy) as defining a kind of entropy
(Tremaine, H\'enon \& Lynden-Bell 1986; Richstone \& Tremaine
1988).
We do not encourage that interpretation,
preferring instead to think of our penalty function
as an {\it ad hoc} numerical device.
See Jaynes (1984) and Binney (1987) for a similar point of view.)
Our goal is to show that there is some choice for $\alpha$ that
reproduces approximately the fully-mixed solutions described above,
while at the same time keeping the orbital weights from varying too
strongly from grid point to grid point.

Figure 13 shows how the deviation from self-consistency, as
measured by $\delta$, depends
on the variance in the orbital weights for the two models
displayed in Figure 12.
We measured the variance via
$$
{\rm VAR}^2(\alpha) = {\sum^M\left(C_i-\overline{C}_i\right)^2\over
M \overline{C}_i^2} \eqno (\new)
$$
with $M$ the number of orbits in the solution and
$\overline{C}_i$ the arithmetic average of the orbital weights.
Because of computer limitations we computed these smooth
solutions using only one-half of the orbits in our libraries; the
$\delta$ values are correspondingly larger than if all the
orbits had been used.
Figure 13 shows exactly the hoped-for behavior: as the solutions are
gradually made smoother (i.e. $\alpha$ is increased above zero),
the departure from self-consistency remains almost constant and
small.
Only when the smoothing parameter $\alpha$ is increased to the
point that the variance becomes unrealistically small ($\ltsim
1$) does $\delta$
become unacceptably large.

{\epsfxsize 5.in \epsfbox{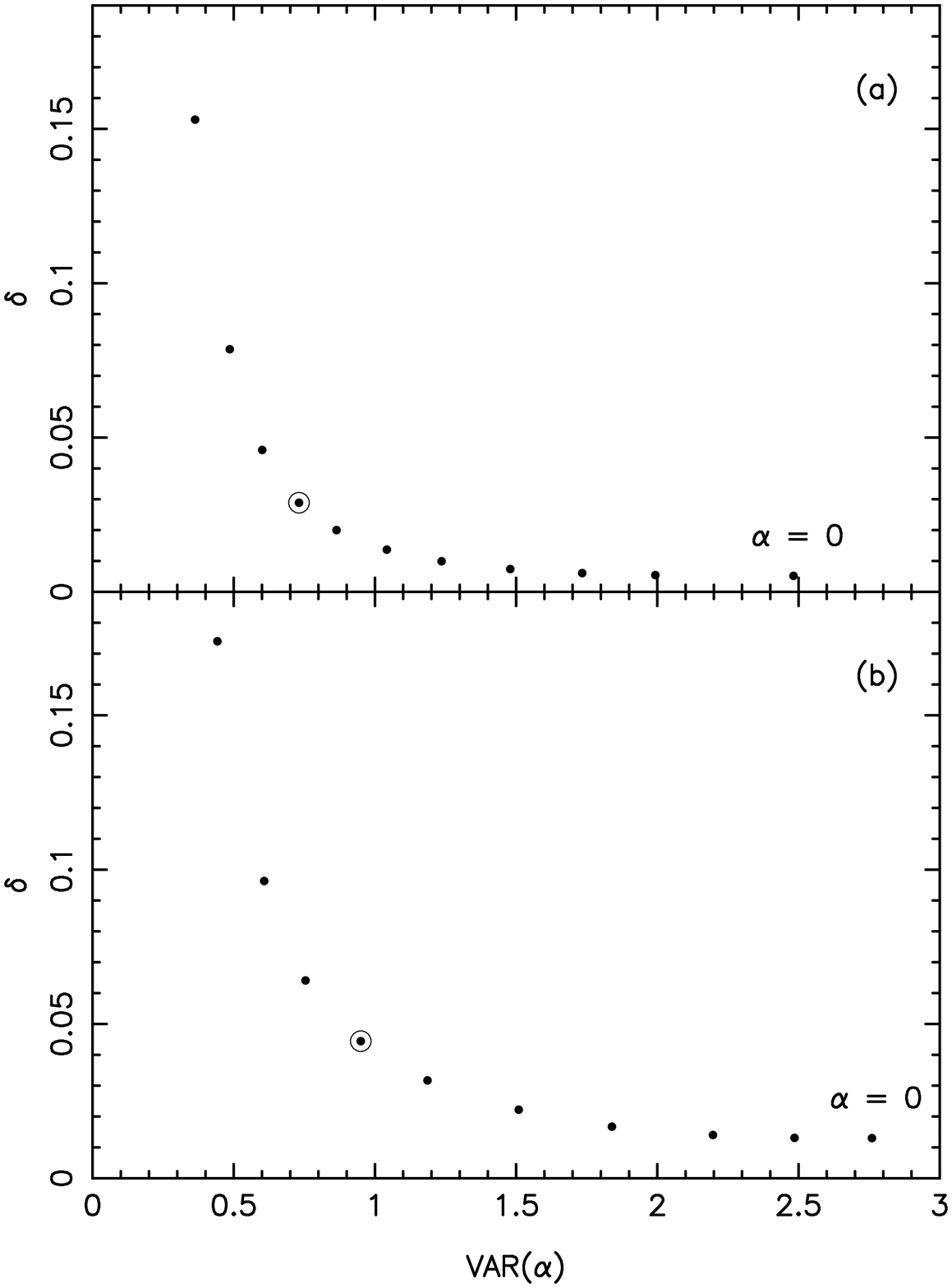}
{\normalbaselineskip=8pt\normalbaselines\noindent
Figure 13. Deviation from self-consistency ($\delta$) as a function of
variance in the orbital weights for the two models illustrated in
Figure 12.
(a) Weak cusp; (b) strong cusp.
The circled dots are the solutions represented in detail in Fig. 14.
\par}

\bigskip

Figure 14 displays the orbital fractions as a function of shell
number for two of these smooth solutions.
The orbital fractions are very similar to those in the unsmooth
solutions, again as hoped; but the dependence of orbital fraction on shell
number is now more nearly continuous, a result of the smoothness
constraint.

These results encourage us to believe that there do exist smooth
solutions corresponding to at least some of our unsmooth ones, and that our
unsmooth solutions have essentially the same, average properties
as the smooth solutions that they are meant to represent.

\sect{Discussion}

This investigation may be seen as a logical extension of the
work of Richstone, Schwarzschild and collaborators
(Richstone 1980, 1982, 1984; Levison \& Richstone 1987; Miralda-Escud\'e
\& Schwarzschild 1989; Lees \& Schwarzschild 1992; Schwarzschild 1993)
on the self-consistency problem for scale-free triaxial galaxies.
Much of that work was directed toward understanding the constraints
that self-consistency places on the shapes of galactic halos.
On kiloparsec scales, the orbital times in galactic halos are
of order 1\% of a Hubble time.
Since stochastic orbits do not behave very differently from regular orbits
over time scales of $\sim 100$ oscillations, Schwarzschild (1993)
adopted the reasonable point of view that stochastic orbits,
integrated for only $\sim 55$ orbital periods, could
safely be included in his solutions without strongly violating the
assumption of stationarity.
He found that stochastic orbits were often required for
self-consistency and showed that their continued evolution
over a Hubble time would not seriously compromise his models.

Although the fraction of chaotic phase space
is not greatly different in our models than
in Schwarzschild's scale-free ones, we have treated the stochastic
orbits in a rather different manner.
Dynamical time scales near the center of an elliptical
galaxy with a cusp are a small fraction of a Hubble time.
As a result, the stochastic orbits can diffuse quickly
through phase space, causing them to lose their
distinguishability in less than a galaxy lifetime.
We found that the number of orbital periods required for stochastic orbits
to visit their allowed phase-space regions in an effectively ergodic manner
is of order $10^2 -- 10^3$ dynamical times in a model with
a $1/r^2$ density cusp,
implying that the stars on stochastic orbits near the centers of
such galaxies should
be distributed with approximately uniform density in their
permitted phase space regions.
This requirement was shown to be inconsistent with dynamical
equilibrium, in the sense that only the innermost 1 or 2 shells of
orbits -- containing $\sim 5\%$ of the mass -- could be constrained
to be fully mixed without violating the self-consistency
equations.
Even in a typical bright elliptical galaxy, where dynamical time
scales are relatively long, one would expect the $\sim20\%$
most-bound stars to have undergone more than $10^3$ orbits during
the lifetime of the galaxy (\S4.2); and this fraction would be larger
in small, dense ellipticals like M32.
Thus we conclude that strong
triaxiality is difficult or impossible to maintain in a galaxy with a
high central concentration of mass.

{\epsfxsize 5.in \epsfbox{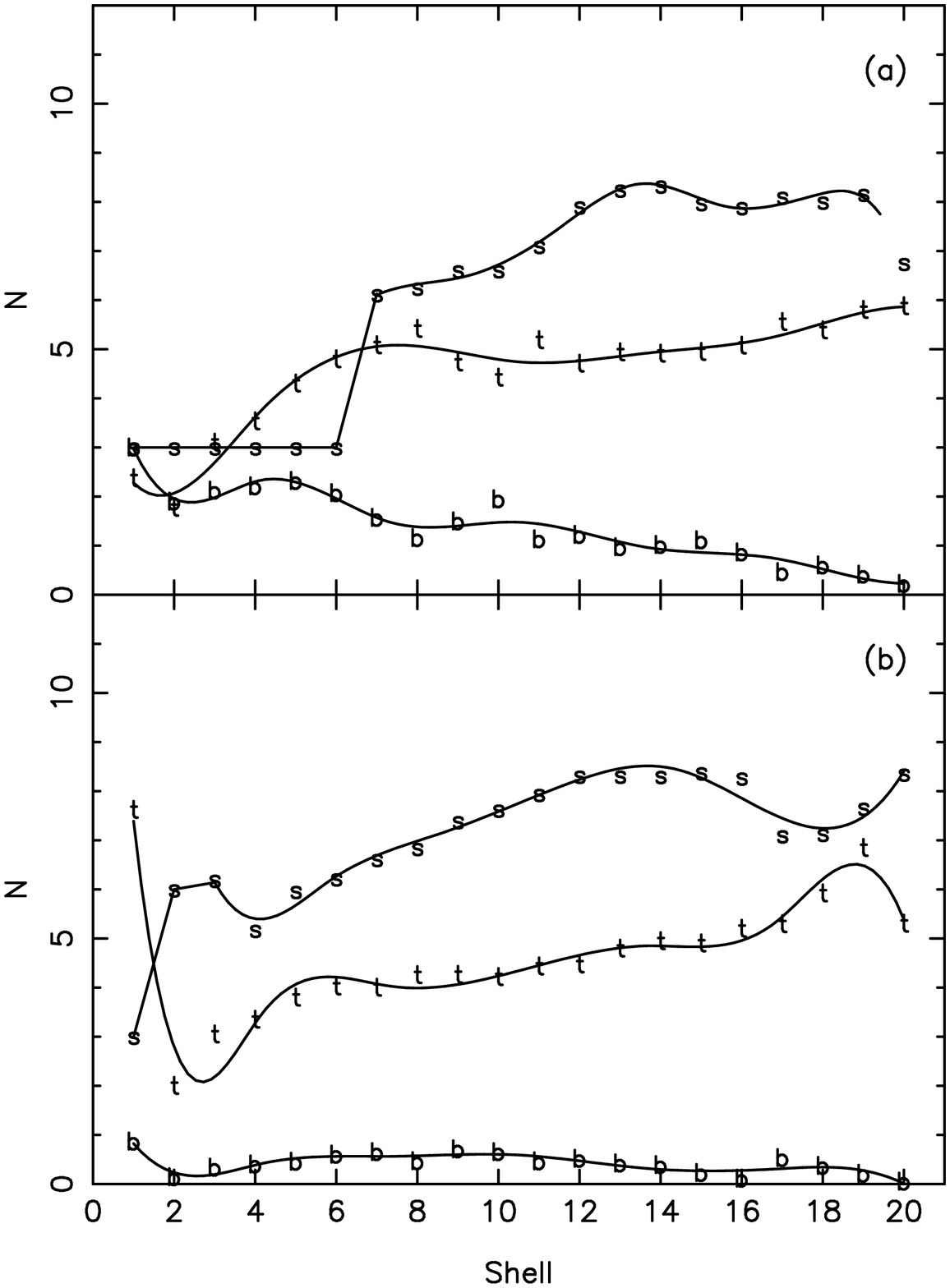}
{\normalbaselineskip=8pt\normalbaselines\noindent
Figure 14. Like Figure 12, for the two smooth solutions indicated in Fig. 13.
(a) Weak cusp; (b) strong cusp.
\par}

\bigskip

In our weak-cusp model, by contrast, a variety of physically
interesting solutions were found, including solutions in which the
stochastic orbits were fully mixed, or simply omitted, within
approximately the half-mass radius.
This greater variety of solutions can be traced to the greater
variety of regular orbit families in the weak-cusp potential,
which permit less weight to be assigned to the stochastic orbits
near the model center.
These models represent galaxies that would evolve very little
over a Hubble time, for two reasons: because the stochastic
parts of phase space are populated in a nearly time-independent
manner at low energies; and because the diffusion time scales for stochastic
orbits in the weak-cusp potential are typically very long, of
order $10^3$ dynamical times or more, so that the non-uniform
population of stochastic phase space at higher energies would not
result in significant evolution over a galaxy lifetime.

Our study raises at least as many questions as it answers.
Among these are:

\medskip
1.{\sl How weak a central cusp can generate significant amounts of
stochasticity in a triaxial potential?}
One of the surprises of this study was the existence of large numbers
of stochastic orbits in triaxial models with density cusps as mild as
$\rho\propto r^{-1}$.
After this study was begun, it was demonstrated that the luminosity densities
of Lauer \etal's (1995) ``core'' galaxies are well described as power
laws, with logarithmic slopes ranging from $-1$ to $0$
(Merritt \& Fridman 1995).
It may be shown that the long-axis orbit, which generates regular
box orbits in integrable or nearly-integrable triaxial
potentials, becomes stable at most energies only when the cusp density
increases more slowly than $r^{-0.5}$ (Merritt \& Fridman 1995);
thus, bona-fide box orbits can only exist in elliptical galaxies
with the weakest observed cusps.
It would be interesting to know whether the fraction of stochastic
phase space in a triaxial potential remains roughly constant as
the cusp slope is reduced to this value,
or whether a larger and larger fraction of the phase space becomes
regular.
Whichever happens, the Liapunov exponents presumably become
smaller as the cusp is made weaker, implying less of a role
for chaos.

In addition to the mass models discussed here, the family:
$$\eqnam{\newmodel}
\rho(m)=\rho_0 (1+m^2)^{-1}(m_0+m^2)^{-1} \eqno (\new)
$$
would be interesting to investigate in this regard.
For $m_0=1$, equation (\newmodel) is the ``perfect'' law (de Zeeuw
1985), while for $m_0=0$ it is similar to the strong-cusp models
presented here.
Thus varying the parameter $m_0$ takes one from a fully integrable
(but non-physical) potential to
a strongly non-integrable (but realistic) potential.

\medskip
2. {\sl How would our results have changed if we had adopted a more
nearly axisymmetric mass model?}
By assuming maximal triaxiality, we forced our solutions to
sample heavily from the non-tube orbits, many of which are
stochastic.
A more nearly oblate or prolate model would have assigned larger
weights to the tube orbits and thus might have achieved
self-consistency without so strongly occupying the stochastic
orbits.
Our results imply only that stochasticity is important in strongly
triaxial, cuspy galaxies.

\medskip
3. {\sl Would the addition of a slow figure rotation strongly
affect the amount of stochasticity in these models?}
Figure rotation is known to induce strong stochasticity at
energies near corotation (e.g. Contopolous 1983).
It might conceivably produce an important change in the degree of
stochasticity at small radii as well, since the axial orbits that
are the generators of stochasticity in our models
become elliptical in the presence of rotation, thus tending to
avoid the central cusp.

\medskip
4. {\sl How would the addition of a central point mass
affect the triaxial self-consistency problem?}
The evidence for central mass concentrations, possibly
supermassive black holes, at the centers of some early-type
galaxies and bulges of spiral galaxies has recently
become very convincing (e.g. Ford \etal~1994; Miyoshi \etal~1995).
Unlike a cusp, a central point mass subjects stars that pass near to
it to large-angle scattering.
Gerhard \& Binney (1985) argued that a central black hole with 2\%
of the ``core'' mass would disrupt most box orbits with apocenters
interior to about 1 kpc.
The phase space of a triaxial model with a mass distribution like
that in galaxy M32 -- which has a strong cusp and an additional, pointlike
mass concentration, possibly a black hole (Tonry 1987) -- would
be even more dominated by chaos than the models examined here.
Significant triaxiality would presumably be very hard to maintain
in such a galaxy.
In fact, axisymmetric models have proved to be very successful at
reproducing the kinematical data for this galaxy
(van der Marel \etal 1994).

\medskip
5. {\sl How accurately can the invariant density associated with
the stochastic part of phase space at a given energy be
approximated by time-averaged trajectories?}
There are at least two issues here, one numerical and one
physical.
A time-averaged regular orbit generates a uniform phase-space
density on its KAM torus; this ``ergodic'' property of regular
orbits justifies the use of time-averaged occupation numbers
when constructing equilibrium models.
No such proof of ergodicity exists for stochastic orbits, and it
is at least possible that even an infinitely long integration of such
an orbit would not produce a uniform filling of the Arnold web.
This possibility is worrisome from the point of view of numerical
construction of equilibrium models, but does not seem desperately
relevant for real galaxies, which care only about the phase-space
distribution at a {\it given} time.
The physically more relevant issue is whether there exist
stochastic trajectories that can remain confined to a limited part of
stochastic phase space even over very long time scales.
Evidence for such ``quasi-regular'' orbits in the weak-cusp
potential was presented in \S 4.2.
Is it fair to populate such orbits with a different density than
that of the other parts of stochastic phase space, or would small
perturbations quickly cause such orbits to sample the entire Arnold
web?

\medskip
6. {\sl Can elliptical galaxies with cusps reach triaxial steady states
like those found here, or would they bypass triaxial configurations in
favor of axisymmetric ones?}
Our work demonstrates the existence of idealized triaxial
models in which the stochastic parts of
phase space are populated in an
approximately uniform, and hence time-independent, manner.
We have not shown that a real galaxy could find these
equilibria.
A real galaxy might always choose axisymmetry or at best only mild
triaxiality, or --- particularly in the case of a galaxy with a weak
cusp --- it might quickly reach a nearly steady, triaxial state and
then continue
evolving slowly in the direction of axisymmetry, from the inside
out.
We do not see any obvious way to choose between these
possibilities based on the modest number of equilibrium models that we have
presented here.
Perhaps the only safe conclusions that we can draw are that chaos
and slow evolution are likely to be generic properties of
triaxial stellar systems.

\bigskip

The computer programs used in this study for integrating orbits,
computing Liapunov exponents and finding and classifying periodic
orbits were written by the Geneva group and graciously
lent to us by S. Udry and D. Pfenniger.
These progams enormously facilitated the work presented here.
Ya. G. Sinai devoted several sessions to explaining ergodic
theory to us and made a number of suggestions that affected the course
of this work.
The universal power-law nature of elliptical galaxy nuclei became
apparent to us early in 1995, after we had been given pre-publication
access to HST surface brightness data by T. Lauer and J. Kormendy.
Conversations with D. Pfenniger and O. Gerhard were invaluable for
understanding the apparent discrepancy described in \S 4 between our
work and theirs.
D. Richstone, J. Sellwood and S. Tremaine
sat through extended presentations of this
work in its late stages and made numerous comments that improved the
final written version.
G. Contopoulos, W. Dehnen, H. Kandrup, N. Murray, G. Quinlan and
M. Schwarzschild also made comments that improved the presentation.
This work was supported by NSF grant AST 90-16515 and by NASA grant NAG
5-2803 to DM.

\vfill\eject

\centerline{Table 1}
\medskip
\centerline{$1:1$ Resonant Orbits in the $x$-$y$ Plane}
\bigskip
\settabs 12 \columns
\hrule\smallskip
\+  & & Radius$^{a}$ & & & & Energy & & & & $T_D$ & \cr
\settabs 7 \columns
\+ Shell & $\gamma=1$ & $\gamma=2$ & $\gamma=1$ &
$\gamma=2$ & $\gamma=1$ & $\gamma=2$ \cr
\hrule
\medskip
\+  1 & 0.2791 & 0.0500 & -0.9964 & -3.789 & 2.444 & 0.160 \cr
\+  2 & 0.4464 & 0.1053 & -0.8670 & -2.884 & 3.381 & 0.342 \cr
\+  3 & 0.6076 & 0.1667 & -0.7695 & -2.359 & 4.262 & 0.551 \cr
\+  4 & 0.7744 & 0.2353 & -0.6887 & -1.989 & 5.176 & 0.790 \cr
\+  5 & 0.9529 & 0.3125 & -0.6186 & -1.704 & 6.167 & 1.068 \cr
\+  6 & 1.148 & 0.4000 & -0.5562 & -1.474 & 7.274 & 1.394 \cr
\+  7 & 1.366 & 0.5000 & -0.4997 & -1.280 & 8.538 & 1.779 \cr
\+  8 & 1.612 & 0.6154 & -0.4478 & -1.114 & 10.01 & 2.240 \cr
\+  9 & 1.896 & 0.7500 & -0.3998 & -0.9695 & 11.76 & 2.800 \cr
\+  10 & 2.226 & 0.9090 & -0.3551 & -0.8412 & 13.88 & 3.489 \cr
\+  11 & 2.620 & 1.100 & -0.3132 & -0.7264 & 16.50 & 4.356 \cr
\+  12 & 3.097 & 1.333 & -0.2738 & -0.6227 & 19.84 & 5.471 \cr
\+  13 & 3.690 & 1.625 & -0.2366 & -0.5285 & 24.21 & 6.945 \cr
\+  14 & 4.449 & 2.000 & -0.2014 & -0.4424 & 30.15 & 8.964 \cr
\+  15 & 5.458 & 2.500 & -0.1680 & -0.3634 & 38.61 & 11.86 \cr
\+  16 & 6.866 & 3.200 & -0.1363 & -0.2905 & 51.44 & 16.28 \cr
\+  17 & 8.974 & 4.250 & -0.1062 & -0.2232 & 72.68 & 23.64 \cr
\+  18 & 12.48 & 6.000 & -0.0776 & -0.1609 & 112.9 & 37.64 \cr
\+  19 & 19.49 & 9.500 & -0.0504 & -0.1031 & 209.2 & 71.26 \cr
\+  20 & 40.49 & 20.00 & -0.0245 & -0.0496 & 596.6 & 207.3\cr
\smallskip
\hrule
\smallskip
$^{a}$ On $x$-axis.

\vfill\eject

\hsize=17.7cm
\vsize=23.9cm

{\offinterlineskip
\centerline{Table 2}
\centerline{Closed orbits in $X-Z$ start space: Model 1 ($\gamma=1$)}}
$$\vbox{
\offinterlineskip
\halign{
\strut \quad # \quad &  \quad # & \quad # &
\quad # & \quad # & \quad # &
 \quad # &  \quad #  \cr \noalign{\hrule}
N &  Resonance & \quad Family & Shell & $x_{0}$ & $y_{0}$ & $z_{0}$ & Stability
\cr \noalign{\hrule}
1 & 1:1:S & xy-loop & 1 & 0.14773 & 0. & 0. & st \cr
  &       &         & 2 & 0.24051 & 0. & 0. & st \cr
  &       &         & 3 & 0.33037 & 0. & 0. & st \cr
  &       &         & 4 & 0.42306 & 0. & 0. & st \cr
  &       &         & 5 & 0.52171 & 0. & 0. & st \cr
  &       &         & 6 & 0.62894 & 0. & 0. & st \cr
  &       &         & 7 & 0.74747 & 0. & 0. & st \cr
  &       &         & 8 & 0.88052 & 0. & 0. & st \cr
  &       &         & 9 & 1.03217 & 0. & 0. & st \cr
  &       &         & 10 & 1.20781 & 0. & 0. & st \cr
  &       &         & 11 & 1.41488 & 0. & 0. & st \cr
  &       &         & 12 & 1.66403 & 0. & 0. & st \cr
  &       &         & 13 & 1.97118 & 0. & 0. & st \cr
  &       &         & 14 & 2.36126 & 0. & 0. & st \cr
  &       &         & 15 & 2.87578 & 0. & 0. & st \cr
  &       &         & 16 & 3.58944 & 0. & 0. & st \cr
  &       &         & 17 & 4.65167 & 0. & 0. & st \cr
  &       &         & 18 & 6.41142 & 0. & 0. & st \cr
  &       &         & 19 & 9.91654 & 0. & 0. & st \cr
  &       &         & 20 & 20.41194 & 0. & 0. & st \cr
\noalign{\hrule} \noalign{\smallskip}
2 & S:1:1 & yz-loop & 1 & 0. & 0. & 0.16602 & st \cr
  &       &         & 2 & 0. & 0. & 0.25916 & st \cr
  &       &         & 3 & 0. & 0. & 0.34662 & st \cr
  &       &         & 4 & 0. & 0. & 0.43549 & st \cr
  &       &         & 5 & 0. & 0. & 0.52941 & st \cr
  &       &         & 6 & 0. & 0. & 0.63121 & st \cr
  &       &         & 7 & 0. & 0. & 0.74378 & st \cr
  &       &         & 8 & 0. & 0. & 0.87042 & st \cr
  &       &         & 9 & 0. & 0. & 1.01526 & st \cr
  &       &         & 10 & 0. & 0. & 1.18375 & st \cr
  &       &         & 11 & 0. & 0. & 1.38342 & st \cr
  &       &         & 12 & 0. & 0. & 1.62499 & st \cr
  &       &         & 13 & 0. & 0. & 1.92448 & st \cr
  &       &         & 14 & 0. & 0. & 2.30701 & st \cr
  &       &         & 15 & 0. & 0. & 2.81433 & st \cr
  &       &         & 16 & 0. & 0. & 3.52158 & st \cr
  &       &         & 17 & 0. & 0. & 4.57891 & st \cr
  &       &         & 18 & 0. & 0. & 6.33655 & st \cr
  &       &         & 19 & 0. & 0. & 9.84490 & st \cr
  &       &         & 20 & 0. & 0. & 20.35492 & st \cr
\noalign{\hrule}
}
} $$

\vfill\eject
{\offinterlineskip
\centerline{Table 3}
\centerline{Closed orbits in stationary start space: Model 1 ($\gamma=1$)}}
$$\vbox{
\offinterlineskip
\halign{
\strut \quad # \quad &  \quad # & \quad # &
\quad # & \quad # & \quad # &
 \quad # &  \quad #  \cr \noalign{\hrule}
N &  Resonance & \quad Family & Shell & $x_{0}$ & $y_{0}$ & $z_{0}$ & Stability
\cr \noalign{\hrule}
3 & 1:2:S & xy-banana & 1 & 0.2791 & 0.0001 & 0. & v \cr
\noalign{\hrule} \noalign{\smallskip}
4 &1:S:2 & xz-banana & 1 & 0.2754 & 0. & 0.0329 & st \cr
  &      &           & 2 & 0.4382 & 0. & 0.0622 & st \cr
  &      &           & 3 & 0.5937 & 0. & 0.0959 & st \cr
  &      &           & 4 & 0.7530 & 0. & 0.1355 & st \cr
  &      &           & 5 & 0.9222 & 0. & 0.1823 & st \cr
  &      &           & 6 & 1.1058 & 0. & 0.2379 & st \cr
  &      &           & 7 & 1.3089 & 0. & 0.3041 & st \cr
  &	 &           & 8 & 1.5371 & 0. & 0.3836 & st \cr
  &      &           & 9 & 1.7977 & 0. & 0.4798 & st \cr
  &      &           & 10 & 2.1003 & 0. & 0.5973 & st \cr
  &      &           & 11 & 2.4581 & 0. & 0.7428 & st \cr
  &      &           & 12 & 2.8900 & 0. & 0.9258 & st \cr
  &      &           & 13 & 3.4243 & 0. & 1.1610 & st \cr
  &      &           & 14 & 4.1054 & 0. & 1.4713 & st \cr
  &      &           & 15 & 5.0069 & 0. & 1.8951 & st \cr
  &      &           & 16 & 6.2613 & 0. & 2.5021 & st \cr
  &      &           & 17 & 8.3135 & 0. & 3.4323 & st \cr
  &      &           & 18 & 11.2411 & 0. & 5.0138 & st \cr
  &      &           & 19 & 17.4355 & 0. & 8.2351 & st \cr
  &      &           & 20 & 35.9663 & 0. & 18.0606 & st \cr \noalign{\hrule}
\noalign{\smallskip}
5  & S:1:2  & yz-banana & 1 & 0. & 0.2484 & 0.0043 & v \cr \noalign{\hrule}
\noalign{\smallskip}
6  & 2:3:S & xy-fish & 1 & 0.2777 & 0.0246 & 0. & v \cr \noalign{\hrule}
\noalign{\smallskip}
7  & 2:S:3 & xz-fish &1 &0.1896 & 0. & 0.0329 & st \cr
 &       &           & 2 & 0.2961 & 0. & 0.2487 & st \cr
 &       &           & 3 & 0.3947 & 0. & 0.3495 & st \cr
 &       &           & 4 & 0.4932 & 0. & 0.4584 & st \cr
 &       &           & 5 & 0.5954 & 0. & 0.5793 & st \cr
 &       &           & 6 & 0.7041 & 0. & 0.7160 & st \cr
 &       &           & 7 & 0.8220 & 0. & 0.8729 & v \cr \noalign{\hrule}
\noalign{\smallskip}
8  & S:2:3 & yz-fish & 1 & 0. & 0.2251 & 0.0865 & v \cr \noalign{\hrule}
\noalign{\smallskip}
9  & 3:4:S & pretzel & 1 & 0.2346 & 0.1343 & 0. & st \cr
 &       &           & 2 & 0.3560 & 0.2404 & 0. & st \cr
 &       &           & 3 & 0.4627 & 0.3536 & 0. & st \cr
 &       &           & 4 & 0.5644 & 0.4784 & 0. & st \cr
 &       &           & 5 & 0.6654 & 0.6183 & 0. & st \cr
 &       &           & 6 & 0.7688 & 0.7767 & 0. & st \cr
 &       &           & 7 & 0.8767 & 0.9583 & 0. & st \cr
 &       &           & 8 & 0.9919 & 1.1683 & 0. & st \cr
 &       &           & 9 & 1.1171 & 1.4142 & 0. & st \cr
}
} $$
$$\vbox{
\offinterlineskip
\halign{
\strut \quad # \quad &  \quad # & \quad # &
\quad # & \quad # & \quad # &
 \quad # &  \quad #  \cr
 &       &           & 10 & 1.2560 & 1.7059 & 0. & v \cr
\noalign{\hrule} \noalign{\smallskip}

10 & 3:4:6 &           & 10 & 1.2691 & 1.6875 & 0.1629 & st \cr
   &     &           & 11 & 1.4444 & 2.0133 & 0.2835 & st \cr
   &     &           & 12 & 1.6509 & 2.4102 & 0.4264 & st \cr
   &     &           & 13 & 1.9006 & 2.9051 & 0.6082 & st \cr
   &     &           & 14 & 2.2124 & 3.5399 & 0.8493 & st \cr
   &     &           & 15 & 2.6173 & 4.3850 & 1.1825 & st \cr
   &     &           & 16 & 3.1716 & 5.5662 & 1.6659 & st \cr
   &     &           & 17 & 3.9864 & 7.3362 & 2.4161 & st \cr
   &     &           & 18 & 5.3205 & 10.2838 & 3.7074 & st \cr
   &     &           & 19 & 7.9492 & 16.1746 & 6.3681 & st \cr
   &     &           & 20 & 15.7302 & 33.8365 & 14.5659 & st \cr
\noalign{\hrule} \noalign{\smallskip}
11 & 4:5:S &           & 1 & 0.1668 & 0.1990 & 0. & v \cr \noalign{\hrule}
\noalign{\smallskip}
12 & 4:5:7 &           & 1 & 0.1669 & 0.1987 & 0.0061 & st \cr
   &     &           & 2 & 0.2459 & 0.3288 & 0.0446 & st \cr
   &     &           & 3 & 0.3124 & 0.4572 & 0.0858 & st \cr
   &     &           & 4 & 0.3737 & 0.5916 & 0.1350 & st \cr
   &     &           & 5 & 0.4330 & 0.7361 & 0.1942 & u \cr \noalign{\hrule}
\noalign{\smallskip}
13 &5:6:S &           & 1 & 0.1126 & 0.2272 & 0.  & st \cr
   &     &           & 2 & 0.1566 & 0.3741 & 0. & st \cr
   &     &           & 3 & 0.1890 & 0.5196 & 0. & st \cr
   &     &           & 4 & 0.2157 & 0.6725 & 0. & st \cr
   &     &           & 5 & 0.2388 & 0.8380 & 0. & v \cr \noalign{\hrule}
\noalign{\smallskip}
14 & 5:6:8 &           & 1 & 0.1229 & 0.1941 & 0.0900 & st \cr
   &     &           & 2 & 0.1756 & 0.3124 & 0.1607 & st \cr
   &     &           & 3 & 0.2172 & 0.4257 & 0.2371 & st \cr
   &     &           & 4 & 0.2537 & 0.5427 & 0.3228 & st \cr
   &     &           & 5 & 0.2873 & 0.6642 & 0.4205 & st \cr
   &     &           & 6 & 0.3196 & 0.7965 & 0.5331 & st \cr
   &     &           & 7 & 0.3515 & 0.9418 & 0.6642 & st \cr
   &     &           & 8 & 0.3843 & 1.1041 & 0.8184 & st \cr
   &     &           & 9 & 0.4184 & 1.2884 & 1.0017 & st \cr
   &     &           & 10 & 0.4554 & 1.5010 & 1.2223 & st \cr
   &     &           & 11 & 0.4963 & 1.7509 & 1.4917 & st \cr
   &     &           & 12 & 0.5433 & 2.0510 & 1.8265 & u \cr
\noalign{\hrule} \noalign{\smallskip}
15 & 5:6:9 &           & 5 & 0.2398 & 0.8372 & 0.0259 & st \cr
   &     &           & 6 & 0.2672 & 1.0141 & 0.0864 & st \cr
   &     &           & 7 & 0.2946 & 1.2111 & 0.1458 & st \cr
   &     &           & 8 & 0.3227 & 1.4337 & 0.2162 & st \cr
   &     &           & 9 & 0.3525 & 1.6887 & 0.3027 & st \cr
   &     &           & 10 & 0.3846 & 1.9857 & 0.4106 & st \cr
   &     &           & 11 & 0.4204 & 2.3376 & 0.5470 & st \cr
   &     &           & 12 & 0.4610 & 2.7629 & 0.7220 & st \cr
}
} $$
$$\vbox{
\offinterlineskip
\halign{
\strut \quad # \quad &  \quad # & \quad # &
\quad # & \quad # & \quad # &
 \quad # &  \quad #  \cr
   &     &           & 13 & 0.5093 & 3.2895 & 0.9507 & st \cr
   &     &           & 14 & 0.5680 & 3.9610 & 1.2568 & u \cr
\noalign{\hrule} \noalign{\smallskip}
16 & 5:7:8 &           & 6 & 0.8225 & 0.1690 & 0.6112 & u \cr
   &     &           & 7 & 0.9495 & 0.2252 & 0.7559 & st \cr
   &     &           & 8 & 1.0867 & 0.2897 & 0.9262 & st \cr
   &     &           & 9 & 1.2378 & 0.3638 & 1.1290 & st \cr
   &     &           & 10 & 1.4075 & 0.4495 & 1.3734 & st \cr
   &     &           & 11 & 1.6022 & 0.5494 & 1.6725 & st \cr
   &     &           & 12 & 1.8313 & 0.6679 & 2.0446 & u \cr
\noalign{\hrule} \noalign{\smallskip}
17 & 6:7:9 &           & 1 & 0.0896 & 0.1687 & 0.1353 & st \cr
   &     &           & 2 & 0.1239 & 0.2685 & 0.2295 & st \cr
   &     &           & 3 & 0.1487 & 0.3621 & 0.3277 & st \cr
   &     &           & 4 & 0.1688 & 0.4559 & 0.4353 & st \cr
   &     &           & 5 & 0.1859 & 0.5533 & 0.5559 & st \cr
   &     &           & 6 & 0.2012 & 0.6570 & 0.6932 & st \cr
   &     &           & 7 & 0.2155 & 0.7694 & 0.8513 & st \cr
   &     &           & 8 & 0.2294 & 0.8937 & 1.0352 & st \cr
   &     &           & 9 & 0.2435 & 1.0336 & 1.2520 & u \cr
\noalign{\hrule} \noalign{\smallskip}
18 &7:5:12 &           & 6 & 0.9421 & 0.1479 & 0.4950 & u \cr
   &     &           & 7 & 1.0969 & 0.2167 & 0.6144 & st \cr
   &     &           & 8 & 1.2648 & 0.3010 & 0.7556 & st \cr
   &     &           & 9 & 1.4498 & 0.4030 & 0.9249 & st \cr
   &     &           & 10 & 1.6571 & 0.5265 & 1.1307 & st \cr
   &     &           & 11 & 1.8939 & 0.6766 & 1.3845 & st \cr
   &     &           & 12 & 2.1705 & 0.8608 & 1.7031 & st \cr
   &     &           & 13 & 2.4943 & 1.0872 & 2.1032 & st \cr
   &     &           & 14 & 2.9119 & 1.3850 & 2.6490 & st \cr
   &     &           & 15 & 3.4416 & 1.7748 & 3.3830 & u \cr \noalign{\hrule}
\noalign{\smallskip}
19 & 7:8:10 &           & 1 & 0.0631 & 0.1425 & 0.1619 & st \cr
    &     &           & 2 & 0.0837 & 0.2240 & 0.2706 & st \cr
    &    &           & 3 & 0.0967 & 0.2980 & 0.3822 & st \cr
    &    &           & 4 & 0.1059 & 0.3701 & 0.5034 & st \cr
    &    &           & 5 & 0.1128 & 0.4428 & 0.6383 & st \cr
    &    &           & 6 & 0.1180 & 0.5182 & 0.7909 & u \cr
\noalign{\hrule} \noalign{\smallskip}
20 & 7:9:10 &           & 1 & 0.1425 & 0.0390 & 0.1735 & st \cr
   &     &           & 2 & 0.2133 & 0.0700 & 0.2881 & st \cr
   &     &           & 3 & 0.2727 & 0.0993 & 0.4053 & st \cr
   &     &           & 4 & 0.3272 & 0.1275 & 0.5323 & st \cr
   &     &           & 5 & 0.3795 & 0.1547 & 0.6735 & st \cr
   &     &           & 6 & 0.4313 & 0.1811 & 0.8330 & st \cr
   &     &           & 7 & 0.4840 & 0.2072 & 1.0155 & st \cr
   &     &           & 8 & 0.5389 & 0.2334 & 1.2269 & st \cr
   &     &           & 9 & 0.5973 & 0.2600 & 1.4747 & u \cr
\noalign{\hrule} \noalign{\smallskip}
}
} $$

\bigskip
{\offinterlineskip
\centerline{Table 4}
\centerline{Closed orbits in $X-Z$ start space: Model 2 ($\gamma=2$)}}
$$\vbox{
\offinterlineskip
\halign{
\strut \quad # \quad &  \quad # & \quad # &
\quad # & \quad # & \quad # &
 \quad # &  \quad #  \cr \noalign{\hrule}
N &  Resonance & \quad Family & Shell & $x_{0}$ & $y_{0}$ & $z_{0}$ & Stability
\cr \noalign{\hrule}
1 & 1:1:S & xy-loop & 1 & 0.02752 & 0. & 0. & st \cr
  &       &         & 2 & 0.05786 & 0. & 0. & st \cr
  &       &         & 3 & 0.09147 & 0. & 0. & st \cr
  &       &         & 4 & 0.12891 & 0. & 0. & st \cr
  &       &         & 5 & 0.17087 & 0. & 0. & st \cr
  &       &         & 6 & 0.21821 & 0. & 0. & st \cr
  &       &         & 7 & 0.27205 & 0. & 0. & st \cr
  &       &         & 8 & 0.33386 & 0. & 0. & st \cr
  &       &         & 9 & 0.40556 & 0. & 0. & st \cr
  &       &         & 10 & 0.48979 & 0. & 0. & st \cr
  &       &         & 11 & 0.59024 & 0. & 0. & st \cr
  &       &         & 12 & 0.71221 & 0. & 0. & st \cr
  &       &         & 13 & 0.86366 & 0. & 0. & st \cr
  &       &         & 14 & 1.05706 & 0. & 0. & st \cr
  &       &         & 15 & 1.31320 & 0. & 0. & st \cr
  &       &         & 16 & 1.66945 & 0. & 0. & st \cr
  &       &         & 17 & 2.20056 & 0. & 0. & st \cr
  &       &         & 18 & 3.08103 & 0. & 0. & st \cr
  &       &         & 19 & 4.83472 & 0. & 0. & st \cr
  &       &         & 20 & 10.08362 & 0. & 0. & st \cr
\noalign{\hrule} \noalign{\smallskip}
2 & S:1:1 & yz-loop & 1 & 0. & 0. & 0.02719 & st \cr
  &       &         & 2 & 0. & 0. & 0.05703 & st \cr
  &       &         & 3 & 0. & 0. & 0.08995 & st \cr
  &       &         & 4 & 0. & 0. & 0.12649 & st \cr
  &       &         & 5 & 0. & 0. & 0.16733 & st \cr
  &       &         & 6 & 0. & 0. & 0.21333 & st \cr
  &       &         & 7 & 0. & 0. & 0.26560 & st \cr
  &       &         & 8 & 0. & 0. & 0.32559 & st \cr
  &       &         & 9 & 0. & 0. & 0.39522 & st \cr
  &       &         & 10 & 0. & 0. & 0.47712 & st \cr
  &       &         & 11 & 0. & 0. & 0.57496 & st \cr
  &       &         & 12 & 0. & 0. & 0.69406 & st \cr
  &       &         & 13 & 0. & 0. & 0.84238 & st \cr
  &       &         & 14 & 0. & 0. & 1.03244 & st \cr
  &       &         & 15 & 0. & 0. & 1.28509 & st \cr
  &       &         & 16 & 0. & 0. & 1.63786 & st \cr
  &       &         & 17 & 0. & 0. & 2.16581 & st \cr
  &       &         & 18 & 0. & 0. & 3.04403 & st \cr
  &       &         & 19 & 0. & 0. & 4.79770 & st \cr
  &       &         & 20 & 0. & 0. & 10.05232 & st \cr
\noalign{\hrule} \noalign{\smallskip}
}
} $$

\vfill\eject
{\offinterlineskip
\centerline{Table 5}
\centerline{Closed orbits in stationary start space: Model 2 ($\gamma=2$)}}
$$\vbox{
\offinterlineskip
\halign{
\strut \quad # \quad &  \quad # & \quad # &
\quad # & \quad # & \quad # &
 \quad # &  \quad #  \cr \noalign{\hrule}
N &  Resonance & \quad Family & Shell & $x_{0}$ & $y_{0}$ & $z_{0}$ & Stability
\cr \noalign{\hrule}
3 & 1:2:S & xy-banana & 1 & 0.0478 & 0.0135 & 0.0000 & v \cr
\noalign{\hrule} \noalign{\smallskip}
4 & 1:S:2 & xz-banana & 1 & 0.0440 & 0. & 0.0186 & st \cr
  &      &           & 2 & 0.0923 & 0. & 0.0396 & st \cr
  &      &           & 3 & 0.1459 & 0. & 0.0636 & st \cr
  &      &           & 4 & 0.2055 & 0. & 0.0911 & st \cr
  &      &           & 5 & 0.2717 & 0. & 0.1224 & st \cr
  &      &           & 6 & 0.3477 & 0. & 0.1596 & st \cr
  &      &           & 7 & 0.4336 & 0. & 0.2027 & st \cr
  &      &           & 8 & 0.5323 & 0. & 0.2535 & st \cr
  &      &           & 9 & 0.6471 & 0. & 0.3142 & st \cr
  &      &           & 10 & 0.7822 & 0. & 0.3876 & st \cr
  &      &           & 11 & 0.9439 & 0. & 0.4775 & st \cr
  &      &           & 12 & 1.1408 & 0. & 0.5899 & st \cr
  &      &           & 13 & 1.3613 & 0. & 0.7681 & st \cr
  &      &           & 14 & 1.7009 & 0. & 0.9216 & st \cr
  &      &           & 15 & 2.1194 & 0. & 1.1775 & st \cr
  &      &           & 16 & 2.7036 & 0. & 1.5429 & st \cr
  &      &           & 17 & 3.5778 & 0. & 2.1012 & st \cr
  &      &           & 18 & 5.0315 & 0. & 3.0481 & st \cr
  &      &           & 19 & 7.9320 & 0. & 4.9734 & st \cr
  &      &           & 20 & 16.6130 & 0. & 10.8368 & st \cr
\noalign{\hrule} \noalign{\smallskip}
5 & S:1:2 & yz-banana & 1 & 0. & 0.0427 & 0.0144 & v \cr
\noalign{\hrule} \noalign{\smallskip}
6 & 2:3:S & xy-fish   & 1 & 0.0358 & 0.0319 & 0. & v \cr
\noalign{\hrule} \noalign{\smallskip}
7 & 2:S:3 & xz-fish   & 1 & 0.0233 & 0. & 0.0353 & v \cr
\noalign{\hrule} \noalign{\smallskip}
8 & S:2:3 & yz-fish   & 1 & 0. & 0.0284 & 0.0314 & v \cr
\noalign{\hrule} \noalign{\smallskip}
9 & 3:4:6 &           & 1 & 0.0264 & 0.0344 & 0.0155 & st \cr
  &      &           & 2 & 0.0549 & 0.0727 & 0.0335 & st \cr
  &      &           & 3 & 0.0856 & 0.1156 & 0.0543 & st \cr
  &      &           & 4 & 0.1190 & 0.1640 & 0.0785 & st \cr
  &      &           & 5 & 0.1556 & 0.2188 & 0.1070 & st \cr
  &      &           & 6 & 0.1959 & 0.2811 & 0.1405 & st \cr
  &      &           & 7 & 0.2407 & 0.3529 & 0.1804 & st  \cr
  &      &           & 8 & 0.2909 & 0.4360 & 0.2283 & st \cr
  &      &           & 9 & 0.3479 & 0.5334 & 0.2863 & st \cr
  &      &           & 10 & 0.4134 & 0.6489 & 0.3574 & st \cr
  &      &           & 11 & 0.4895 & 0.7874 & 0.4455 & st \cr
  &      &           & 12 & 0.5807 & 0.9580 & 0.5578 & st \cr
  &      &           & 13 & 0.6913 & 1.1710 & 0.7026 & st \cr
  &      &           & 14 & 0.8300 & 1.4448 & 0.8951 & st \cr
}
} $$
$$\vbox{
\offinterlineskip
\halign{
\strut \quad # \quad &  \quad # & \quad # &
\quad # & \quad # & \quad # &
 \quad # &  \quad #  \cr
  &      &           & 15 & 1.0099 & 1.8101 & 1.1600 & st \cr
  &      &           & 16 & 1.2558 & 2.3210 & 1.5424 & st \cr
  &      &           & 17 & 1.6160 & 3.0862 & 2.1330 & st \cr
  &      &           & 18 & 2.2031 & 4.3588 & 3.1445 & st \cr
  &      &           & 19 & 3.3525 & 6.8965 & 5.2190 & st \cr
  &      &           & 20 & 6.7336 & 14.4866 & 11.5849 & u \cr
\noalign{\hrule} \noalign{\smallskip}
10 & 4:5:7 &           & 1 & 0.0131 & 0.0345 & 0.0238 & st \cr
   &     &           & 2 & 0.0268 & 0.0728 & 0.0512 & st \cr
   &     &           & 3 & 0.0410 & 0.1149 & 0.0827 & st \cr
   &     &           & 4 & 0.0559 & 0.1619 & 0.1190 & st \cr
   &     &           & 5 & 0.0720 & 0.2144 & 0.1613 & st \cr
   &     &           & 6 & 0.0890 & 0.2737 & 0.2106 & st \cr
   &     &           & 7 & 0.1074 & 0.3410 & 0.2688 & st \cr
   &     &           & 8 & 0.1274 & 0.4183 & 0.3379 & st \cr
   &     &           & 9 & 0.1494 & 0.5078 & 0.4209 & st \cr
   &     &           & 10 & 0.1739 & 0.6130 & 0.5215 & st \cr
   &     &           & 11 & 0.2015 & 0.7383 & 0.6455 & u \cr
\noalign{\hrule} \noalign{\smallskip}
11 & 4:6:7 &           & 1 & 0.0325 & 0.0106 & 0.0286 & st \cr
   &     &           & 2 & 0.0677 & 0.0225 & 0.0612 & st \cr
   &     &           & 3 & 0.1058 & 0.0359 & 0.0984 & st \cr
   &     &           & 4 & 0.1474 & 0.0512 & 0.1414 & st \cr
   &     &           & 5 & 0.1929 & 0.0685 & 0.1910 & u \cr
\noalign{\hrule} \noalign{\smallskip}
12 & 5:7:8 &           & 1 & 0.0253 & 0.0105 & 0.0331 & st \cr
   &     &           & 2 & 0.0522 & 0.0220 & 0.0706 & st \cr
   &     &           & 3 & 0.0812 & 0.0346 & 0.1136 & st \cr
   &     &           & 4 & 0.1126 & 0.0483 & 0.1628 & st \cr
   &     &           & 5 & 0.1467 & 0.0636 & 0.2195 & st \cr
   &     &           & 6 & 0.1840 & 0.0805 & 0.2852 & st \cr
   &     &           & 7 & 0.2252 & 0.0994 & 0.3621 & st \cr
   &     &           & 8 & 0.2713 & 0.1206 & 0.4527 & st \cr
   &     &           & 9 & 0.3234 & 0.1446 & 0.5606 & st \cr
   &     &           & 10 & 0.3831 & 0.1723 & 0.6906 & st \cr
   &     &           & 11 & 0.4527 & 0.2045 & 0.8495 & u \cr
\noalign{\hrule} \noalign{\smallskip}
}
} $$
\hsize=16cm
\vsize=21cm

\vfill\eject

{\refs

Aarseth, S. J. \& Binney, J. J. 1978, \MNRAS 185, 227

Arnold, V. I. 1964, Russian Math. Surveys 18, 85

Benettin, G., Galgani, L., Giorgilli, A. \& Strelcyn, J.-M. 1980,
Meccanica 15, 21

Binney, J. J. 1982a, \MNRAS 201, 1

Binney, J. J. 1982b, \MNRAS 201, 15

Binney, J. J. 1987, in IAU Symposium No. 127, Structure and
Dynamics of Elliptical Galaxies, ed. T. de Zeeuw (Dordrecht:
Reidel), p. 269

Carollo, C. M. 1993, Ph.D. thesis, Ludwig-Maximilians Univ.,
Munich

Chandrasekhar, S. 1969, Ellipsoidal Figures of Equilibrium (New
York: Dover), p. 52

Chirikov, B. V. 1979, Phys. Rept. 52, 263

Contopoulos, G. 1983, \AAp 117, 89

Contopoulos, G. \& Barbanis, B. 1989, \AAp 222, 329

Crane, P. {\it et al.} 1993, \AJ 106, 1371

Crone, M., Evrard, A. E. \& Richstone, D. O. 1994, \ApJ 434, 402

Dehnen, W. 1993, \MNRAS 265, 250

de Zeeuw, P. T. 1985, \MNRAS 216, 273

de Zeeuw, P. T. \& Lynden-Bell, D. 1985, \MNRAS 215, 713

de Zeeuw, P. T. \& Pfenniger, D. 1988, \MNRAS 235, 949

Fehlberg, E. 1968, NASA Technical Report TR R-287

Ferrarese, L, van den Bosch, F. C., Ford, H. C., Jaffe, W., \&
O'Connell, R. W. 1994, \AJ 108, 1598

Ford, H. C. \etal 1994, \ApJL 435, L27

Gebhardt, K. {\it et al.} 1996, preprint.

Gebhardt, K., Richstone, D. O. {\etal} 1996, preprint

Gerhard, O. E. 1987, in IAU Symposium No. 127, Structure and
Dynamics of Elliptical Galaxies, ed. T. de Zeeuw (Dordrecht:
Reidel), p. 241

Gerhard, O. E. \& Binney, J. J. 1985, \MNRAS 216, 467

Goodman, J. \& Schwarzschild, M. 1981, \ApJ 245, 1087

Heggie, D. 1991, in Predictability, Stability and Chaos in $N$-Body
Dynamical Systems, ed. A. E. Roy (New York: Plenum), p. 47.

Henon, M. \& Heiles, C 1964, \AJ 69, 73

Hernquist, L. 1990, \ApJ 356, 359

Jaffe, W. 1983, \MNRAS 202, 995

Jaffe, W., Ford, H. C., O'Connell, R. W. van den Bosch, F. C. \&
Ferrarese, L. 1994, \AJ 108, 1567

Jaynes, E. T. 1984, in Inverse Problems, ed. D. W. McLaughlin
(Providence: SIAM), p. 151

Kormendy, J. 1985, \ApJ 292, L9

Kormendy, J., Dressler, A., Byun, Y.-I., Faber, S. M.,
Grillmair, C., Lauer, T. R., Richstone, D., \& Tremaine, S. 1995,
in ESO/OHP Workshop on Dwarf Galaxies,
ed. G. Meylan \& P. Prugniel (Garching: ESO), p. 147

Kuijken, K. 1993, \ApJ 409, 68

Kuzmin, G. G. 1973, in The Dynamics of Galaxies and Star Clusters, ed.
T. B. Omarov (Nauka of the Kazakh S. S. R., Alma-Ata), p. 71.

Lauer, T. {\it et al.} 1995, preprint.

Lees, J. F. \& Schwarzschild, M. 1992, \ApJ 384, 491

Levison, H. F. \& Richstone, D. O. 1987, \ApJ 314, 476

Lichtenberg, A. J. \& Lieberman, M. A. 1992, Regular and Chaotic
Dynamics (New York: Springer)

Lynden-Bell, D. 1967, \MNRAS 136, 101

Merritt, D. 1996, {\it Cel. Mech.}, in press

Merritt, D. \& Fridman, T. 1995, in Fresh Views of Elliptical
Galaxies, ed. A. Buzzoni, A. Renzini \& A. Serrano, in press

Miller, G.F. 1974, in Numerical Solutions of Integral Equations,
ed. L.M. Delves \& J. Walsh (Oxford: Clarendon), p.175

Miralda-Escud\'e, J. \& Schwarzschild, M. 1989, \ApJ 339, 752

Miyoshi, M. \etal 1995, Nature 373, 127

Moller, P., Stiavelli, M. \& Zeilinger, W. W. 1995, \MNRAS 276,
979

Nieto, J.-L. \& Bender, R. 1989, \AAp 215, 266

O'Sullivan, F. 1986, {\it Stat. Sci.} 1, 502

Pfenniger, D. \& de Zeeuw, T. 1989, in Dynamics of Dense Stellar
Systems, ed. D. Merritt (Cambridge: Cambridge), p. 81

Phillips, D. L. 1962, J. Ass. Comput. Mach. 9, 84

Richstone, D. O. 1980, \ApJ 238, 103

Richstone, D. O. 1982, \ApJ 252, 496

Richstone, D. O. 1984, \ApJ 281, 100

Richstone, D. O. \& Tremaine, S. 1988, \ApJ 327, 82

Schwarzschild, M. 1979, \ApJ 232, 236

Schwarzschild, M. 1982, \ApJ 263, 599

Schwarzschild, M. 1993, \ApJ 409, 563

Sinai, Ya. G. 1976, Introduction to Ergodic Theory (Princeton:
Princeton University Press)

Statler, T. S. 1987, \ApJ 321, 113

Tikhonov, A. N. 1963, Soviet Math. 4, 1035

Tonry, J. L. 1987, \ApJ 322, 632

Tremaine, S., Richstone, D., Buyn, Y.-I., Dressler, A., Faber, S.
M., Grillmair, C., Kormendy, J., \& Lauer, T. R. 1994, \AJ 107, 634

Tremaine, S., Henon, M. \& Lynden-Bell, D. 1986, \MNRAS 219, 285

Udry, S. \& Pfenniger, D. 1988, \AAp 198, 135

van der Marel, R. P., Evans, N. W., Rix, H. W., White, S. \& de Zeeuw, T.
1994, \MNRAS 271, 99

Wilkinson, A. \& James, R. A. 1982, \MNRAS 199, 171

}

\end